\documentclass[preprint,3p,times,authoryear]{elsarticle}
\usepackage{amssymb}
\usepackage{comment}
\usepackage{mathpazo}
\usepackage{eulervm}
\usepackage[utf8]{inputenc}

\newcommand\Theader{\rule[-1.5ex]{0pt}{5ex}}

\usepackage[table,xcdraw]{xcolor}
\usepackage{pgfplotstable}

\usepackage{balance}  % to better equalize the last page
\usepackage{graphics} % for EPS, load graphicx instead
\usepackage[T1]{fontenc}
\usepackage{mathptmx}
\usepackage[pdftex]{hyperref}
\usepackage{soul}
\usepackage{booktabs}
\usepackage{multicol}
\usepackage{enumitem}
\usepackage{float}
\usepackage{url}
\usepackage{parskip}
\usepackage{comment}
\usepackage{microtype} % Improved Tracking and Kerning
\usepackage{ccicons}  % Cite your images correctly!
\usepackage{array}
\newcolumntype{P}[1]{>{\centering\arraybackslash}p{#1}}
\usepackage{todonotes}
\usepackage{relsize,etoolbox}%

\usepackage{color}
\usepackage{colortbl}
\usepackage{multirow}

  {\list{}{\leftmargin=0.3in\rightmargin=0.3in}\item[]}%
  {\endlist}

\newlength{\Oldarrayrulewidth}
\definecolor{intnull}{RGB}{213,229,255}
\definecolor{inteins}{RGB}{128,179,255}
\definecolor{intzwei}{RGB}{42,127,255}
\definecolor{intdrei}{RGB}{0,85,212}
\definecolor{intvier}{RGB}{0,51,128}
\definecolor{intfunf}{RGB}{0,34,85}

\definecolor{niceBlue}{RGB}{42,68,109}
\pgfplotstableset{
    color cells/.style={
        col sep=comma,
        string type,
        postproc cell content/.code={%
                \pgfkeysalso{@cell content=\rule{0cm}{2.4ex}\cellcolor{niceBlue!##1}\pgfmathtruncatemacro\number{##1}\ifnum\number>50\color{white}\fi##1}%
                },
        columns/x/.style={
            column name={}, column type={r},
            postproc cell content/.code={}
        }
    }
}

\usepackage{tabularx}

\journal{Elsevier}
\usepackage{relsize,etoolbox}
\AtBeginEnvironment{quote}{\smaller}

\hypersetup{draft}
\begin{document}

\begin{frontmatter}

%% Title, authors and addresses

%% use the tnoteref command within \title for footnotes;
%% use the tnotetext command for theassociated footnote;
%% use the fnref command within \author or \address for footnotes;
%% use the fntext command for theassociated footnote;
%% use the corref command within \author for corresponding author footnotes;
%% use the cortext command for theassociated footnote;
%% use the ead command for the email address,
%% and the form \ead[url] for the home page:
%% \title{Title\tnoteref{label1}}
%% \tnotetext[label1]{}
%% \author{Name\corref{cor1}\fnref{label2}}
%% \ead{email address}
%% \ead[url]{home page}
%% \fntext[label2]{}
%% \cortext[cor1]{}
%% \address{Address\fnref{label3}}
%% \fntext[label3]{}

\title{Everything you always wanted to know about a dataset: studies in data summarisation}

%% use optional labels to link authors explicitly to addresses:
%% \author[label1,label2]{}
%% \address[label1]{}
%% \address[label2]{}

%\author{}

%\address{}
\address[label1]{University of Southampton, UK}
\address[label2]{The Open Data Institute, UK}

\author[label1,label2]{Laura Koesten\corref{corrauthor}}
\cortext[corrauthor]{The corresponding author can be contacted via email at laura.koesten@theodi.org and is based at The Open Data Institute, 65 Clifton Street, London EC2A 4JE Tel: +44 20 3598 9395}
%\ead{emilia.kacprzak@theodi.org}

%\orcid{1234-5678-9012-3456}
%\address{%
%  \institution{University of Southampton, The Open Data Institute, UK}
%  \city{London}
%  \country{United Kingdom}
%}

\author[label1]{Elena Simperl}
%\address{%
%  \institution{University of Southampton, UK}
%  \city{Southampton}
%  \country{United Kingdom}
%}

\author[label1,label2]{Emilia Kacprzak}
%\address{%
%  \institution{University of Southampton, The Open Data Institute, UK}
%  \city{London}
%  \country{United Kingdom}
%}

\author[label1]{Tom Blount}
%\address{%
% \institution{University of Southampton, UK}
% \city{Southampton}
% \country{United Kingdom}
%}

\author[label2]{Jeni Tennison}
%\address{%
%  \institution{The Open Data Institute, UK}
%  \city{London}
%  \country{United Kingdom}
%}

\begin{abstract}
Summarising data as text helps people make sense of it. It also improves data discovery, as search algorithms can match this text against keyword queries. In this paper, we explore the characteristics of text summaries of data in order to understand how meaningful summaries look like. We present two complementary studies: a data-search diary study with $69$ students, which offers insight into the information needs of people searching for data; and a summarisation study, with a lab and a crowdsourcing component with overall $80$ data-literate participants, which produced summaries for $25$ datasets. In each study we carried out a qualitative analysis to identify key themes and commonly mentioned dataset attributes, which people consider when searching and making sense of data. The results helped us design a template to create more meaningful textual representations of data, alongside guidelines for improving data-search experience overall.
\end{abstract}

\begin{keyword}
data summarisation \sep data search \sep data sensemaking \sep human data interaction
%% PACS codes here, in the form: \PACS code \sep code

%% MSC codes here, in the form: \MSC code \sep code
%% or \MSC[2008] code \sep code (2000 is the default)

\end{keyword}

\end{frontmatter}

%% \linenumbers

\section{Introduction}
\label{sec:introduction}
As digital technology has advanced over the past years, there has been a huge surge in data. Structured and semi-structured data in particular, which refers to data that is organised explicitly, for example as spreadsheets, web tables, databases and maps, has become critical in most domains and professional roles \citep{McKinsey:report}. With the rise of data science, millions of such datasets have been published, sometimes under an open license, in institutional repositories, online marketplaces and on social networks in sectors from science and finance to marketing and government \citep{DBLP:journals/corr/abs-1801-04971,Verhulst2016}.\footnote{In this paper, a 'dataset' refers to structured or semi-structured information collected by an individual or organisation, which is distributed in a standard format, for instance as CSV files. In the context of search, it refers to the artifacts returned by a search algorithm in response to a user query.} Recently, Google has announced an initiative to use its schema.org markup language to index datasets alongside text documents, images and products.\footnote{\url{http://schema.org/Dataset}}

Previous research has shown that, despite increased availability, this data cannot be easily reused, as people still experience many difficulties in finding, accessing and assessing it \citep{Laura}. In \citet{Laura} we discussed three major aspects that matter to data practitioners when selecting a dataset to work with: \textit{relevance}, \textit{usability} and \textit{quality}. For each of these aspects, people have to make sense of the content and context of a dataset to make an informed decision about whether to use it for their task. This process demonstrates unique interaction characteristics, which have been subject to several human data interaction studies \citep{DBLP:journals/corr/GregoryGCSW17,DBLP:conf/ercimdl/KernM15,Laura, Boukhelifa:2017:DWC:3025453.3025738}.

\begin{figure*}
\centering
%\begin{footnotesize}
\includegraphics[width=160mm,scale=0.5]{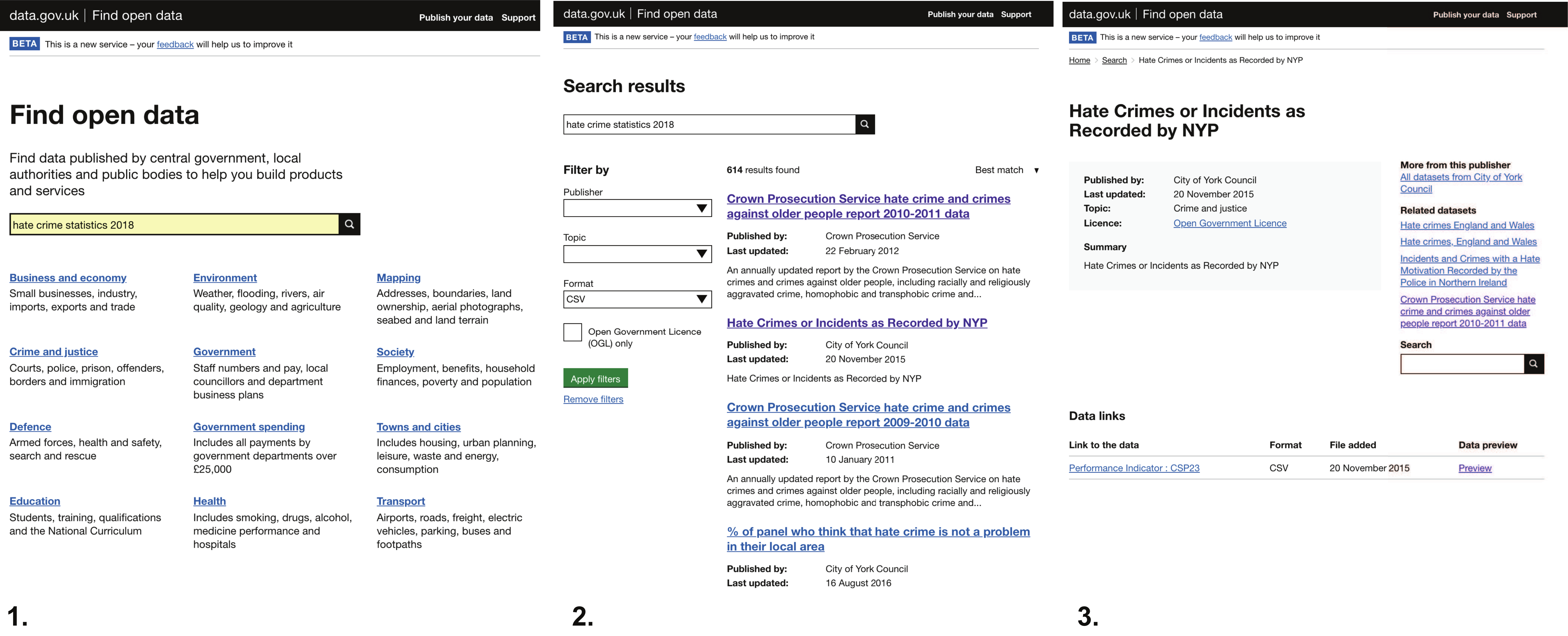}
%\end{footnotesize}
\caption{Data search on a data portal}
\label{fig:searchprocess}
\end{figure*}

Data search often starts on a data portal with an interface as depicted in Figure \ref{fig:searchprocess}. Upon entering their query, users are presented with a compact representation of the results, which includes for each dataset its metadata (title, publisher, publication date, format etc.), a short snippet of text, and, in some cases, a data preview or a visualisation. Figure \ref{fig:searchresult} shows an example. From a user's perspective, having a textual summary of the data is paramount: text is easier to digest than raw data or graphs \citep{law2005comparison,van2010graph}, and richer in context than metadata. It helps people assess the relevance, usability and quality of a dataset for their own needs~\citep{DBLP:conf/adcs/AuTJ16,bargmeyer2000metadata,Lehmberg2016,nguyen2015result}. It also improves data discovery, as search algorithms can match the text against keyword queries~\citep{DBLP:conf/adcs/ThomasOR15}.

In general, a good summary must be able to represent the core idea, and effectively convey the meaning of the source \citep{zhuge2015dimensionality}. In this paper, we aim to understand what this means in a data context: what a dataset summary must capture in order to help data practitioners select the data to work with with more confidence. This is currently a gap in the human data interaction literature. The data engineering community, on the other hand, has created some standards and best practices for publishing and sharing data, including DCAT,\footnote{\url{https://www.w3.org/TR/vocab-dcat/}} schema.org,\footnote{\url{http://schema.org/Dataset}} and SharePSI)\footnote{\url{https://www.w3.org/2013/share-psi/bp/}}). However, none of these initiatives offer any guidance on what to include in a dataset summary. Sometimes text summaries are generated automatically using so-called natural language generation (NLG) methods. These methods are commonly bootstrapped via parallel corpora of data and text snippets, but an extensive exploration of the qualities of these training corpora is missing \citep{DBLP:journals/corr/WisemanSR17}. Overall, this leads to summaries that vary greatly in terms of content, language and level of detail, which are often not fit for purpose~\citep{Laura,DBLP:journals/jdiq/NeumaierUP16}.

We have undertaken two complementary studies in dataset selection and summarisation. Both studies build on previous research of ours from \citet{Laura}. In that work, we have reported on the results of a series of interviews with $20$ data practitioners, which have helped us define a general \textit{framework for dataset search}, built around the three themes mentioned earlier: \textit{relevance}, \textit{usability} and \textit{quality}. The first study presented in this paper takes the next step: we analysed $69$ data-search diaries by students who were asked to document in detail how they go about finding and selecting datasets. The students wrote $269$ diaries, which we analysed qualitatively starting from the framework from \citep{Laura}. This resulted in a \textit{list of dataset selection attributes}. In the second study, we carried out lab and crowdsourcing experiments with overall $80$ data-literate participants, who created a total of $360$ summaries for $25$ datasets. We analysed the summaries thematically to derive common structures in their composition, which led to a \textit{list of dataset summary attributes}. We grouped these attributes into four main types of information: (i) \textit{basic metadata} such as format and descriptive statistics; (ii) \textit{dataset content}, including major topic categories, as well as geospatial and temporal aspects; (iii) \textit{quality statements}, including uncertainty; and (iv) \textit{analyses and usage ideas}, such as trends observed in the data.

We found a core set of attributes that were consistently prevalent in the two studies, across different datasets and participants. We used them to define a \textit{template} to design more meaningful textual representations of data, which resonate with what people consider relevant when describing a dataset to others, and when trying to make sense of a dataset they have not used before.

Our summary template is primarily meant as a tool for data publishers, but also for data scientists and engineers. It could be integrated into data publication forms alongside common metadata fields. It could also help build data-to-text algorithms that do a better job at reflecting the information needs and expectations of summary readers, and improve dataset indexing strategies, which are currently relying on metadata \citep{DBLP:conf/compsac/ReicheH13, marienfeld2013metadata}. The findings of the two studies also suggest much needed extensions to existing metadata standards in order to cover aspects such as the numbers of rows and columns in a dataset; the levels of granularity of temporal and geospatial information; quality assessments; and meaningful groupings of headers.

\newpage
\noindent \textbf{Summary of research questions and contributions.}
Our paper explores the following research questions:

\begin{description}
\item[RQ1] Study $1$: What data attributes do people consider when determining the relevance, usability and quality of a dataset?
\item[RQ2] Study $2$: What data attributes do people choose to mention when summarising a dataset to others?
\end{description}

Our paper contributes to the emerging field of human data interaction by presenting, to the best of our knowledge, the first in-depth characterisation of human-generated dataset summaries. The two studies helped us identify, on the one hand, dataset attributes which people find useful to make sense of a dataset, and, on the other hand, attributes they choose to describe a dataset to others. Both informed the design of the summary template. Our aim was to create practical, user-centric guidelines for data summarisation that reflect the needs and expectations of data consumers rather than what data publishers consider important. The work expands our understanding of how people interact with and communicate about data, and can further inform the design of data publishing platforms, metadata standards,  and algorithms for natural language generation and dataset retrieval.

%\TODO{change first figure and align it with search query from Section 2. For Figure 2, highlight the summary. }
%\setkeys{Gin}{draft}

%Moreover, a better understanding of dataset summaries for sensemaking has potential implications when structured data is used in decision making contexts. Helping people to find data when they need it allows them to make decisions more effectively. \TODO{T:read}

%when an overview of a set of data can be crucial \cite{gkatzia2016natural1}.
%For instance, summaries are used as decision making support in medical setting to understand specific patient records \cite{law2005comparison1}. They allow to highlight particular patterns and trends and link them to events which can improve clinical decision making \cite{DBLP:journals/aicom/GattPRHMMS09,scott2013data}.
%understand the condition of a patient, as has been shown in medical decision-making
%n. Being able to quickly understand the content of a dataset can help....

%based on what people search for and find important about it. \TODO{repeated} We are interested in what people choose to communicate about a dataset that is new to them - what they see and how they decide to write about it. This gives us insights in how people perceive and make sense of data that they find.\TODO{why are we interested in this?}

\section{Motivating scenario}
\label{sec:motivatingscenario}
Before describing the two studies and their context, we will expand on the data search example introduced earlier to give an overview of the state of the art, and of the challenges that motivate our work. Imagine you want to analyse trends in street crime rates in London over the past year. You are trying to find data that is relevant for this information need/task. An overview of the process is depicted in Figure \ref{fig:searchprocess}.

\begin{figure}[H]
\centering
\includegraphics[width=75mm]{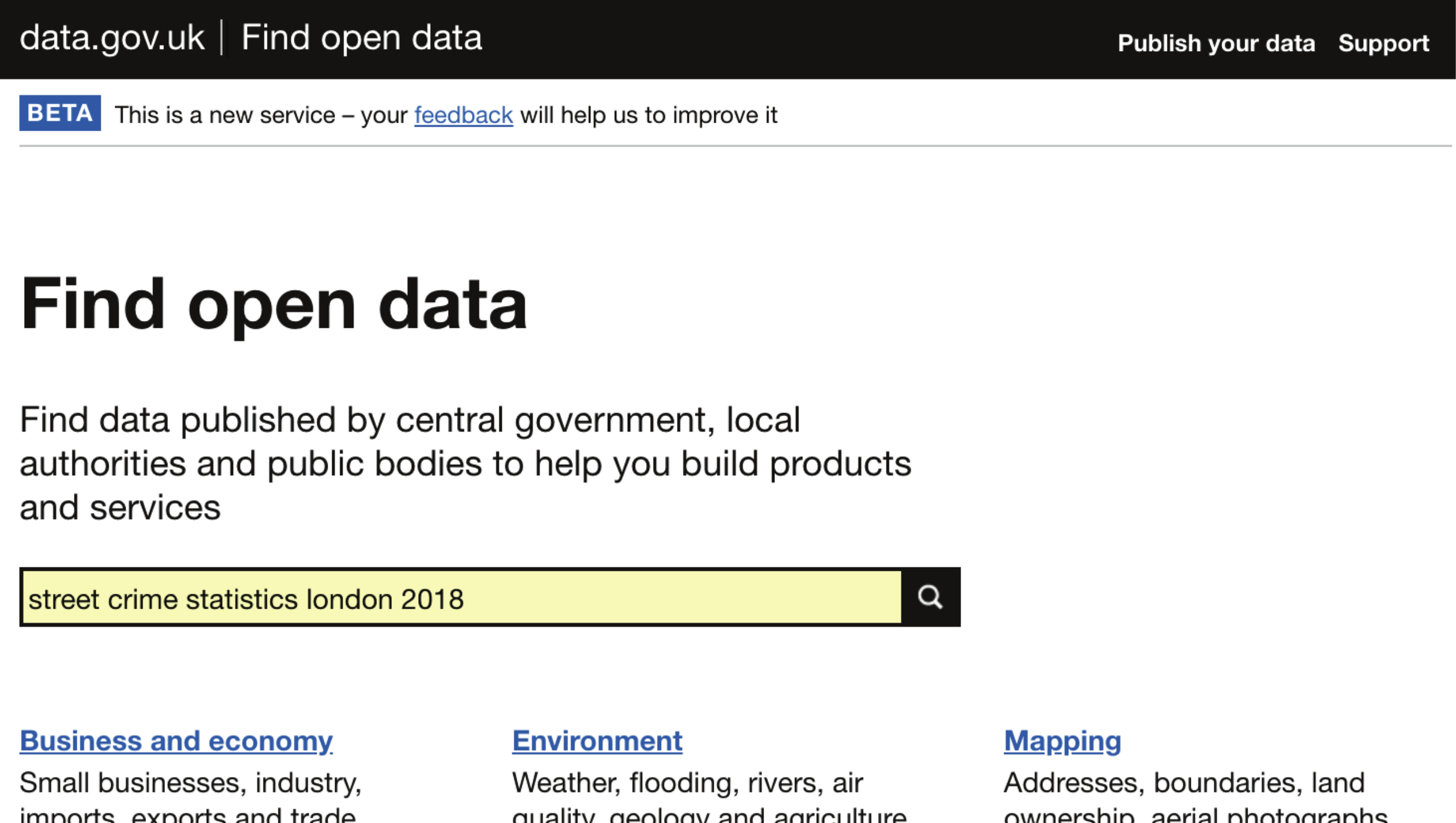}
\caption{Data search interface on a data portal. Users can type keywords in the search field or browse the data collection by domain and other attributes (1)}
\label{fig:portalinterface}
%\vspace{-3mm}
\end{figure}

You enter a search query such as \textit{'hate crime statistics 2018'} in the search box of a UK data portal (see Figure \ref{fig:portalinterface}).

%\setkeys{Gin}{draft}
%\setkeys{Gin}{draft=false}

\begin{figure}[H]
\centering
\begin{minipage}{.49\textwidth}
\centering
\includegraphics[width=62mm, scale=0.5]{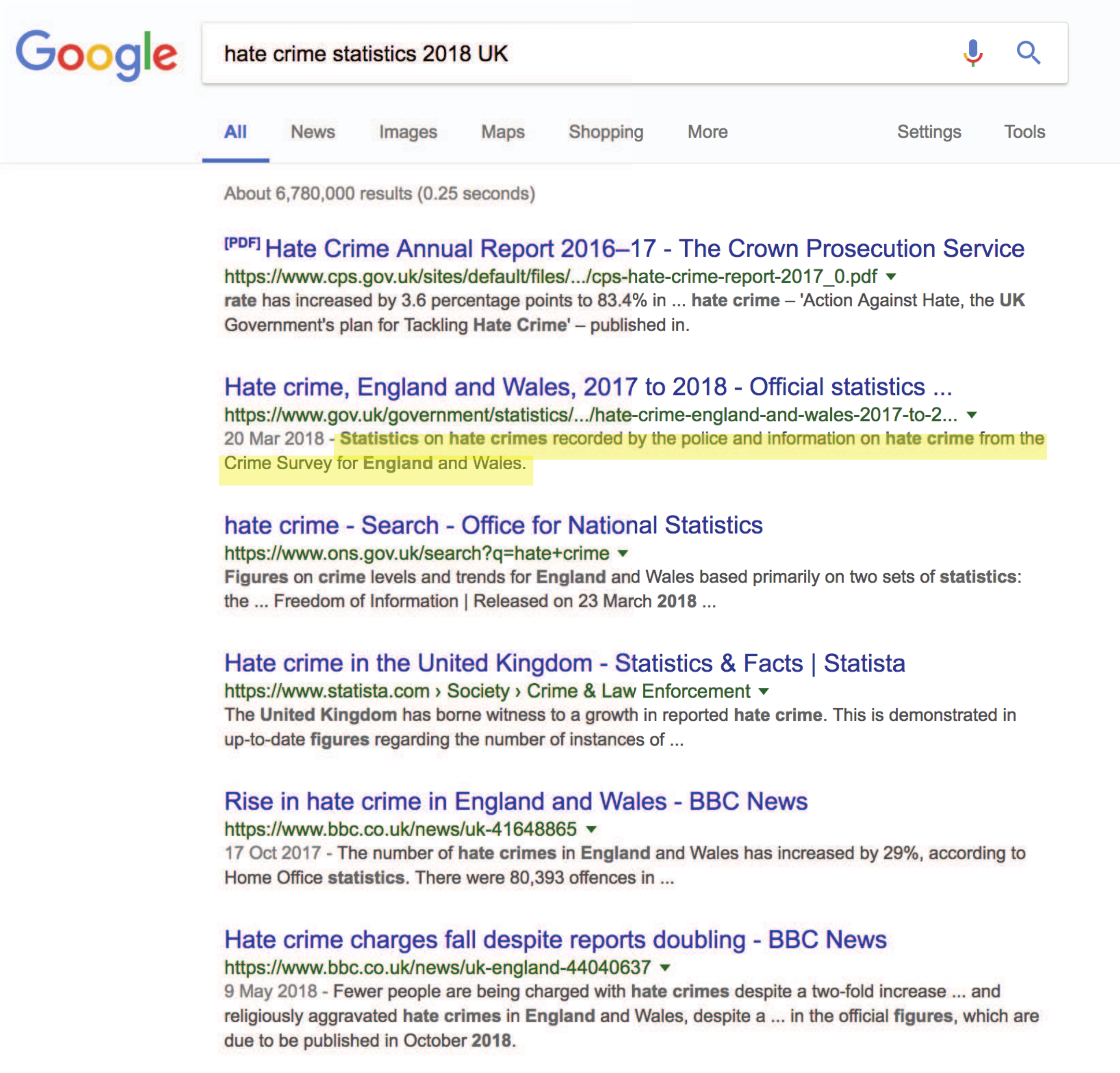}
\caption{Example of a data search result list on Google. \newline Underneath each search result is a a snippet describing \newline the underlying website which may or may not contain data.}
\label{fig:googleresults}
\end{minipage}
\begin{minipage}{.49\textwidth}
\centering
\includegraphics[width=55mm, scale=0.5]{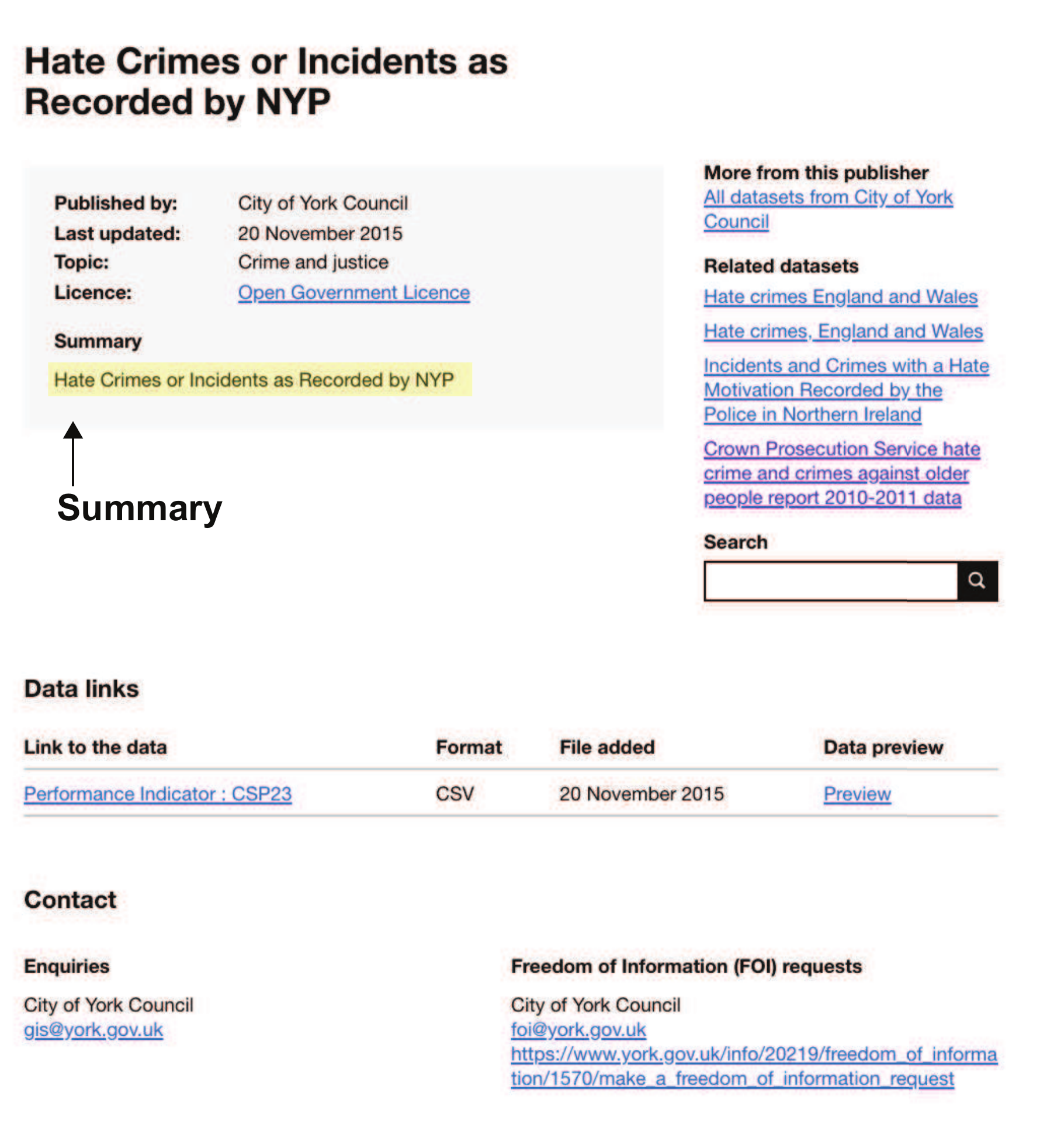}
\caption{A dataset preview page on the open data portal \newline of the UK government}
\label{fig:metadata}
\end{minipage}
\end{figure}

\begin{figure*}[h]
\centering
%\begin{footnotesize}
\includegraphics[width=100mm]{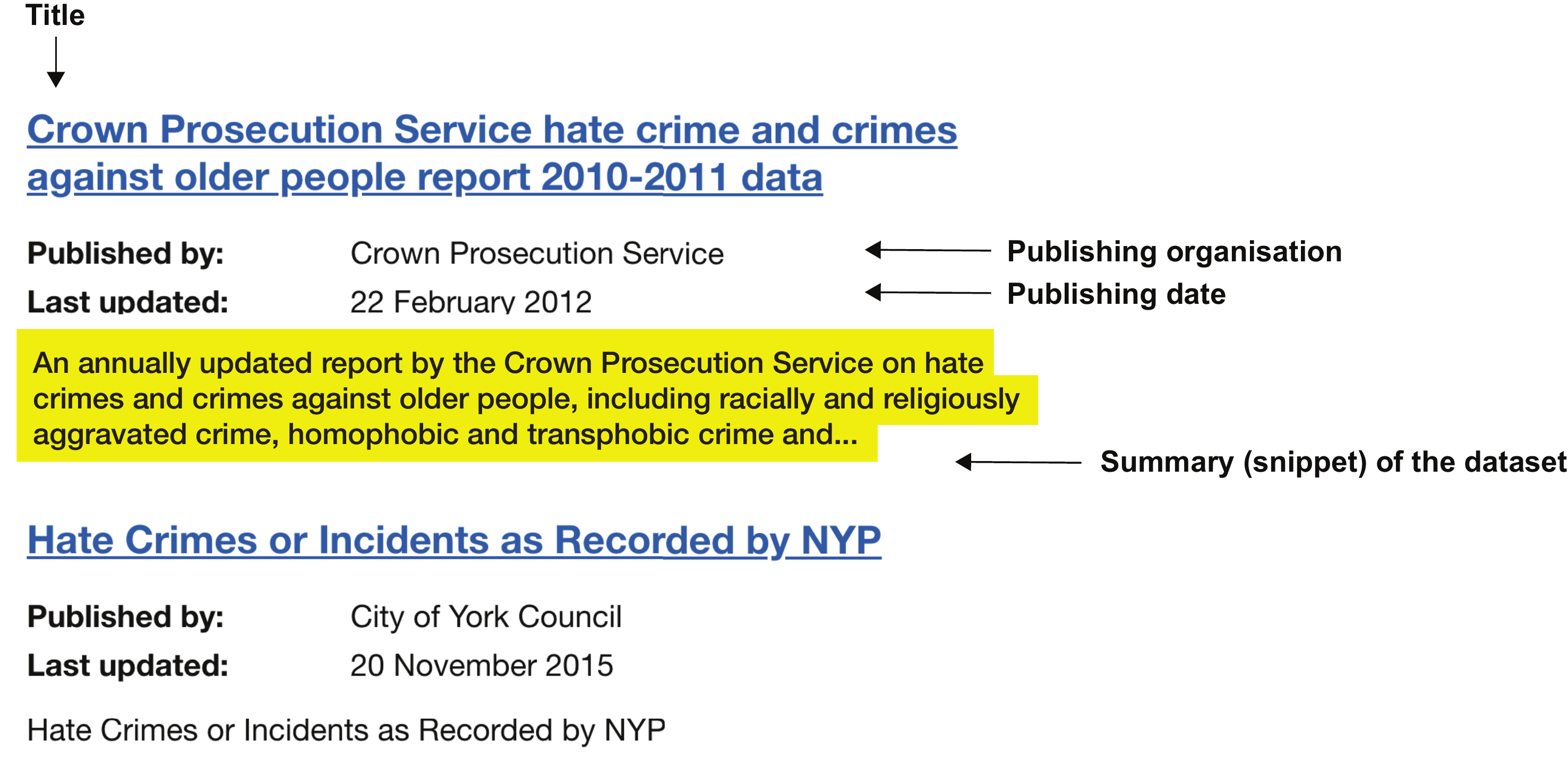}
%\end{footnotesize}
%\vspace{-7mm}
\caption{Example of a data search result list on one of the most popular open government data portals. Next to title, publisher, domain and format, we see a textual description of the dataset}
\label{fig:searchresult}
%\vspace{-5mm}
\end{figure*}

The search results may look like in Figure \ref{fig:searchresult}. On the left hand side, you can find a classification of the results based on metadata attributes such as license and format, as well as the number of datasets that fall into each category. You can use these facets to explore the collection of datasets or filter the results. On the right hand side, you can choose from a ranked list of datasets. Each dataset is presented via its metadata with title, publisher, main domain and available formats. In most cases, the dataset is accompanied by a short text summary. The results can be sorted according to different criteria, including relevance. You select one of the results to explore further, based on what else is on the list and on the information displayed in the snippet. This commonly takes you to a new page (see Figure \ref{fig:metadata}), where you can also download the dataset to examine it on your own computer.

%\setkeys{Gin}{draft}

Data search results on general-purpose web search engine have a similar look and feel (see Figure \ref{fig:googleresults}), although the hits are a mix of datasets and other types of sources. In this case, you might be able to tell from the result snippets which links refer to datasets, click on the results, and look for a download link, a table or an API.

\begin{comment}
%put together with other figure
\begin{figure}
\centering
\begin{footnotesize}
\includegraphics[width=80mm]{figures/googleW}
\end{footnotesize}
%\vspace{-7mm}
\caption{Example of a data search result list on Google. Underneath each search result is a a snippet describing the underlying website which may or may not contain data.}
\label{fig:googleresults}
%\vspace{-5mm}
\end{figure}
%\setkeys{Gin}{draft=false}
\end{comment}
No matter where the search journey starts, a textual description is often key to determine whether a dataset is fit-for-purpose, or if you need to continue the search. Our two studies aim to understand the characteristics of this crucial element in the interaction between people and data.

\begin{comment}
\begin{figure}[h]
%\centering
\begin{footnotesize}
\includegraphics[width=85mm]{figures/crimedatagov}
\end{footnotesize}
%\vspace{-3mm}
\caption{Example search result list for street crime in London on the UK national open data portal. Each result has a more detailed preview page which contains a download link}
\label{fig:crimedatagov}
%\vspace{-2mm}
\end{figure}
\end{comment}
%\TODO{Add dataset screenshot and mark the summary}

\begin{comment}
%\setkeys{Gin}{draft}
\begin{figure}[H]
\centering
\begin{footnotesize}
\includegraphics[width=80mm]{figures/metadatagovuk}
\end{footnotesize}
%\vspace{-5mm}
\caption{A dataset preview page on the open data portal of the UK government}
\label{fig:metadata}
\end{figure}
%\setkeys{Gin}{draft=false}
\end{comment}

\section{Related work}
\label{sec:relatedwork}
In this section we outline current practices for selecting and describing data on the web and related work on summarisation. We draw on metadata standards and community guidelines for data publishing; literature in the fields of data search and sensemaking, as well as  natural language generation; and related HCI and HDI (human data interaction) studies.  %\TODO{changesentence}

\subsection{Selecting and making sense of datasets}
\label{sec:relatedworkselection}

A rich body of information retrieval literature explores how people select documents and determine their relevance to a given task or information need \citep{barry1994user,park1993nature, schamber1990re}. We also know that different information sources result in people searching and choosing results differently, as relevance depends on context. This has been shown in search verticals, for instance for scientific publications \citep{Li2010,Yu:2005:ATD:1092358.1092535}, people~\citep{Weerkamp2011peoplesearch} or products \citep{DBLP:conf/cikm/GyselRK16,rowley2000product}.

Previous works have highlighted the distinct characteristics of dataset compared to document retrieval. Data requires context to create meaning and make sense of it. While this may apply to search in general, choosing a dataset greatly depends on the information provided alongside it, for example via metadata or graphs. For example, \citet{DBLP:conf/jcdl/WynholdsWBST12} show that, in the context of digital libraries, seeking and using documents and data for research purposes are different in terms of information needs, processes and required level of support. In user studies with social scientists, \citep{DBLP:conf/ercimdl/KernM15} found that the quantity and quality of metadata is more critical in dataset search than in literature search, where convenience prevails. Empirical social scientists in that study were willing to put more effort into the retrieval of research data than in literature retrieval. In our prior mixed-methods study mentioned earlier \citep{Laura} we found that specific relevance, usability and quality aspects were perceived to be different for data than for documents - for example, the methodology used to collect and clean the data, missing values, the granularity of the captured information, as well as the ability to understand the schema used to organise a dataset and to process it in the form it was published.

% \begin{figure}[h]
% \begin{footnotesize}
% \includegraphics[width=78mm]{figures/framework}
% \end{footnotesize}
% \vspace{-5mm}
% \caption{Framework for Human Interaction with structured data, describing the information seeking process for structured data on the web, resulting from an interview study with data professionals and a search log analysis of a large open data portal. The analysis of data searching diaries in this work focuses on the Evaluation task of the selection process.}
% \label{fig:framework}
% \end{figure}

A review of related literature \citep{DBLP:journals/jis/BalatsoukasMO09} concluded that textual metadata surrogates, if designed in a user-centred way, can help people identify relevant documents and increase accuracy and/or satisfaction with their relevance judgements. Several authors have shown that textual summaries perform better in decision making than graphs. For instance, \citet{DBLP:journals/aicom/GattPRHMMS09} found in a evaluation of a system that summarises patient data that all users (doctors and nurses) perform better in decision making tasks after viewing a text summary with manually generated text versus a graph. These findings are confirmed by \citep{law2005comparison,van2010graph} in studies comparing textual and graphical descriptions of physiological data displayed to medical staff. \citet{DBLP:conf/chi/SultanumBWC18} emphasise the need to integrate textual summaries to get an overview of clinical documentation instead of relying on graphical representations.

As a starting point for the studies presented in this paper, we thus made two assumptions: (i) textual summaries for datasets can be written by people without having an in-depth knowledge of data analysis and visualisation techniques; and (ii) summaries help data practitioners decide whether to use a dataset or not with more confidence. While we do not claim that text-based surface representations are superior to graphs, we believe they are, at a minimum, complementary to visualisations and accessible to a broad range of audiences, including less experienced users~\citep{DBLP:journals/aicom/GattPRHMMS09,Laura, DBLP:conf/chi/SultanumBWC18}. Our findings support our assumptions. The crowdsourcing experiments showed that summaries can be created by people with basic data literacy skills who are not familiar with the dataset. In addition, across the two studies reported here we were able to identify common themes and attributes of summaries that match the information needs of potential readers.

For the remainder of this section, we will elaborate on existing practices and techniques to create text about structured data.

\subsection{Practices around text summaries for datasets}

When searching on the web, we are used to being presented with a snippet, which is the short summarising text component that is returned by a search engine for each hit. This helps us make a decision about the relevance of the documents returned \citep{Bando:2010:CQS:1840784.1840813}. Snippets adjust their content based on the user query to make selection more effective \citep{Bando:2010:CQS:1840784.1840813}. There are initial efforts that aim to do the same for dataset search \citep{DBLP:conf/adcs/AuTJ16}, but we are still very far from being able to provide the same user experience as in web search.

Currently dataset summaries are created by people, often the data publishers, who might take metadata standards and community guidelines as a point of reference. Existing community guidelines for data sharing, such as the W3C's Data on the Web Best Practices\footnote{\url{https://www.w3.org/TR/dwbp/}} or SharePSI focus on the machine readability of data. Textual descriptions are part of the standards, but guidelines for what should they contain are sparse.

This can be seen, for instance, in the W3C's Data on the Web Best Practices, which is based on DCAT, a vocabulary to describe datasets in catalogues, or, in a slightly different context, in the documentation of schema.org, a set of schemas for structured data markup on web pages. Instructions are formulated as follows:
\begin{quote}
%\vspace{-4px}
(DCAT) \textit{description = free-text account of the dataset (rdfs:Literal)} \\ (schema.org) \textit{description = a description of the item (text)}.
\end{quote}

Based on the general lack of guidance, we focus the current paper on understanding the composition of meaningful summaries rather than exploring whether people find the resulting summaries useful, which we believe is a necessary next step. There are very few studies that empirically evaluate any of the existing metadata standards in user studies - most efforts so far have concentrated on providing guidance for those who add information to a dataset, in many cases the data publishers. For the purpose of consistency, in this paper we refer to textual descriptions of datasets as \textit{summaries}.

\subsection{Human-generated summaries of datasets}

Summarising text is a complex and well-studied area of research in domains such as education, linguistics and psychology, amongst others \citep{yu2009shifting}. The cognitive processes triggered by this task, as studied in psychology, are described as involving three distinct activities: (i) \textit{selection} (selecting which aspects of the source should be included in the summary); (ii) \textit{condensation} (substitution of source material through higher-level ideas, or more specific lower-level concepts); and (iii) \textit{transformation} (integrating and combining ideas from the source) \citep{Bando:2010:CQS:1840784.1840813}.

Johnson defines a summary as a brief statement that represents the condensation of information accessible to a subject and reflects the central ideas or essence of the discourse \citep{hidi1986producing}. Describing or summarising something is a language activity and based in culture: the concepts, definitions and understandings developed in a community. Differences in cultural contexts can lead to misinterpretation of dataset content or to difficulties in developing a common understanding of a dataset summary. Constructing meaning from information - in our case the dataset and the accompanying summary  - is always constructed by the reader, and is influenced by a variety of confounding factors.

Literature on text summarisation differentiates between writer-based summaries, which are summaries written for the writer herself, and reader-based summaries, which are written for an audience and usually require some planning \citep{hidi1986producing}. In this paper we consider the latter.

Our research, as much of the related work in human data interaction, is based on the assumption that, in order to offer the best user experience, we cannot simply reuse or re-purpose principles and models that have been proposed for less structured sources of information \citep{marchionini2005accessing,wilson2010keyword}. Summarising structured or semi-structured data is inherently different to summarising free text. The complexity of constructing meaning from structured data (in contrast to text) has been discussed in the literature \citep{marchionini1997information,DBLP:conf/avi/PirolliR96}. Understanding data requires cognitive work in order to contextualise it in relation to other information, and context to make it meaningful \citep{albers2015human}; arguably more than when summarising natural language text \citep{DBLP:journals/corr/Gkatzia16}.

In their review of summary generation from text, \citep{DBLP:journals/air/GambhirG17} point out the subjectivity of the task and the lack of objective criteria for what is important in a summary. Summary quality is suggested to depend on its purpose, focus and particular requirements of the task \citep{Owczarzak:2009:EAS:1708155.1708161}. In our studies we follow a similar view - we compare themes and datasets attributes derived from the summaries created in our lab and crowdsourcing experiments with analogue attributes prevalent to the task context in which such summaries would be most likely used, which is dataset selection and sensemaking.

\subsection{Automatic summary generation}
Automatically summarising text to accurately and concisely capture key content is an established area of research, with a wide range of techniques, most recently neural networks, employing language models of varying degrees of sophistication \citep{boydell2007social,DBLP:journals/air/GambhirG17,reiter1997building,DBLP:journals/corr/WisemanSR17}.
Two broad approaches to summarisation are reported in the literature: (i) \textit{extraction} (or intrinsic summarisation); and (ii) \textit{abstraction} (or extrinsic summarisation). Extractive approaches aim to create a summary by selecting content from the source they are summarising. Abstractive approaches aim to paraphrase the original source to provide a higher-level content representation \citep{boydell2007social}. Research has focused more on extractive methods as abstractive methods are rather complex \citep{DBLP:journals/air/GambhirG17}. Based on existing studies in human data interaction, we believe that meaningful dataset summaries likely require abstractive elements including quality statements, descriptive statistics or topical coverage of a dataset \citep{DBLP:journals/corr/GregoryGCSW17,DBLP:conf/ercimdl/KernM15,Laura, Boukhelifa:2017:DWC:3025453.3025738}.
%\TODO{check references at the end of the previous sentence}

Automatic generation of summaries for data is a comparatively newer field, although there have been significant advances in this area \citep{DBLP:journals/corr/WisemanSR17}. Current approaches tend to mostly work in closed domains and the complexity of performing these tasks is acknowledged in literature \citep{DBLP:journals/corr/MeiBW15a}.
Data-to-text generation has been explored in several areas, such as health informatics \citep{gatt2009data,scott2013data}, weather forecasts \cite{gkatzia2016natural,sripada2004lessons},
finance \citep{Kukich:1983:DKR:981311.981340}, sports reporting \citep{DBLP:journals/corr/WisemanSR17}; as well as for different data formats, such as in graphs, databases and trend series \citep{Bechchi:2007:MDD:1321440.1321500,Cormode:2015:CSO:2745754.2745781,liu2014distributed,roddick1999methods,sripada2003summarizing,yu2007choosing}. Recognised subtasks in this space include: content selection (selecting what data gets used in the summary) and surface realisation (how to generate natural language text about the selected content) \citep{gkatzia2016content}.

Summaries produced with data-to-text generation methods are at the moment usually extractive rather than abstractive and tend to be merely textual representations of the dataset content, almost like a textual 'visualisation' (e.g.\citep{DBLP:journals/corr/WisemanSR17}):

\begin{quote}
\textit{Extract taken from an automatically generated summary from Wiseman et al., 2017: }
The Atlanta Hawks defeated the Miami Heat
, 103 - 95 , at Philips Arena on Wednesday.
Atlanta was in desperate need of a win and
they were able to take care of a shorthanded
Miami team here. Defense was key for
the Hawks , as they held the Heat to 42
percent shooting and forced them to commit
16 turnovers. Atlanta also dominated in the
paint, winning the rebounding battle, 47
- 34, and outscoring them in the paint 58
- 26. The Hawks shot 49 percent from the
field and assisted on 27 of their 43 made
baskets. %This was a near wire - to - wire win for the Hawks , as Miami held just one lead in the first five minutes. Miami ( 7 - 15 ) are as beat - up as anyone right now and it ’s taking a toll on the heavily used starters . Hassan Whiteside really struggled in this game , as he amassed eight points , 12 rebounds and one blocks on 4 - of - 12 shooting...
\end{quote}

Our work helps define commonly used strategies for abstracting the content of a dataset in a summarisation context; as such, it can inform the design of abstractive approaches by pointing to types of information that an algorithm should aim to include in a summary to make it more useful for its readers.

\begin{comment}
\begin{figure}[H]
\centering
%\vspace{-3mm}
\begin{tiny}
\includegraphics[width=55mm]{figures/wiseman}
\end{tiny}
\caption{Extract taken from an automatically generated summary from Wiseman et al., 2017}
\label{fig:wiseman}
%\vspace{-3mm}
\end{figure}
\end{comment}

Another closely related area of research is data profiling, which refers to a wide range of methods to describe datasets, with a focus on their numerical or structural properties. Profiles can be merely descriptive or include analysis elements of a dataset \citep{Naumann:2014:DPR:2590989.2590995}. Some approaches connect the dataset to other resources to add more context or to generate richer profiles, for example spatial or topical profiles \citep{DBLP:conf/esws/FetahuDNCTN14,shekhar2010spatial}. Most papers in this space work on datasets from a specific domain or on particular types of data such as graphs or databases. The result is not necessarily a human-readable text summary, but a reduced, higher-level version of the original dataset \citep{DBLP:conf/vldb/Saint-PaulRM05}.

Much of the work in automatic summary generation require gold standards for evaluation  \citep{Bando:2010:CQS:1840784.1840813}. These corpora are typically created manually, but their quality is uncertain and guidelines and best practices are largely missing \citep{DBLP:journals/air/GambhirG17}. Summary evaluation covers metrics computed automatically (e.g. BLEU Rouge, etc.), human judgement or a combination of the two \citep{DBLP:journals/air/GambhirG17,Owczarzak:2009:EAS:1708155.1708161}. A deep understanding of the best ways to run human evaluations, which criteria to use, the biases they create and so on on is not available - most studies use criteria such as accuracy, readability, coverage etc. but they are small-scale and not analysed in great detail. We believe this is partially due to a limited appreciation of what a meaningful summary should contain. Evidence for best-practice dataset summaries could lead to more meaningful evaluation methodologies in this space, by informing the design of evaluation benchmarks.

\section{Methodology}
\label{sec:methodology}
\begin{figure}[H]
\centering
\begin{footnotesize}
\includegraphics[width=75mm]{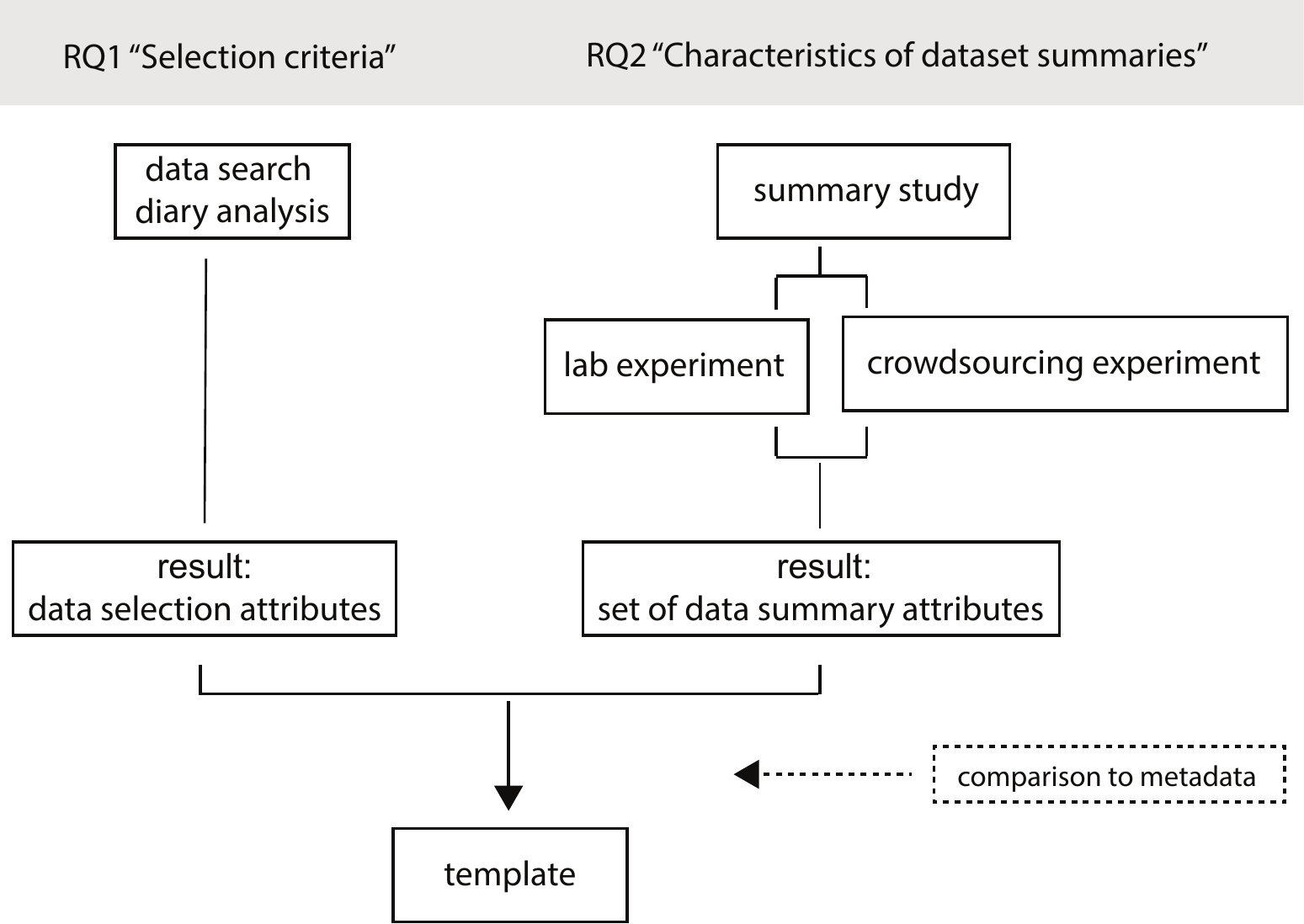}
\end{footnotesize}
\caption{Overview of research methods and outcomes}
\label{fig:methodologyoverview}
%\vspace{-3mm}
\end{figure}

Figure \ref{fig:methodologyoverview} gives an overview of the research carried out. To answer \textbf{RQ1} we conducted a thematic analysis of data-search diaries which resulted in a \textit{list of data selection attributes}. For \textbf{RQ2} we applied a mixed-methods approach \citep{bryman2006integrating} combining a task-based lab experiment and a crowdsourcing experiment, in which participants summarised datasets in a writing task. This led to a \textit{list of data summary attributes}. Across the studies, we were able to identify core attributes that were prevalent for different datasets and participants. We compared them to existing metadata standards for data publication and sharing to understand existing gaps and design a summary template.

\subsection{Study $1$: Data-search diaries}
\label{sec:diarystudy}
In the first study we analysed data-search diaries to get an in-depth understanding of the criteria that influence people's decisions to choose a dataset to work with. This also gave us insight into the kinds of information that need to be captured in dataset summaries to make them more useful for dataset sensemaking and selection.

\noindent \textbf{Process.} We conducted a thematic analysis of $269$ data-search diaries that were completed by $69$ students\footnote{MSc Data Science ($n$=49), MSc Computer Science ($n$=10), MSc Operational Research and Finance ($n$=6), MEng Computer Science ($n$=3), MSc Operational Research \& Statistics ($n$=1)} for a data science project within a university course. The participants were actively searching for datasets to work with and were instructed to write a diary entry for each data search task for two weeks. They were asked to find two to five datasets for their coursework. They were free to choose the topic of their project -  there was hence no domain restriction to the datasets they could use or to the way they searched for the data.

The students were encouraged to document their data seeking behaviour directly after each search session and to reflect upon their data selection choices. The overall aim was to make them aware of the range of factors that come into play when looking for data, and of the importance of data sourcing for data science work.

We provided an online form with open-ended diary questions. The students self-selected when and what to report. For the purpose of our study, we focused on a subset of diary questions that concerned selection criteria for datasets:
\vspace{5px}
\begin{itemize}[noitemsep]
\scriptsize
\item What do you need to know about a dataset before you select it for your task?
\item What is most important for you when selecting a dataset for this task?
\item What tells you that the data is useful and relevant for your task?
\item What tells you that the data is good quality for your task?
\end{itemize}
\vspace{5px}
\noindent \textbf{Analysis.}
The free-text answers to these questions were analysed using thematic analysis \citep{robson2016real}. Two of the authors deductively coded the answers based on the framework for human structured-data interaction from \citep{Laura}, which defines \textit{relevance}, \textit{usability} and \textit{quality} as general themes in dataset selection. As a second layer of coding we open-coded attributes emerging in each of these areas. In this step, the coding was done by one researcher, but to enhance reliability two senior researchers checked the analysis for a sample of the data. The analysis resulted in a \textit{list of data selection attributes}. As noted earlier, they helped us understand what kinds of information good summaries need to contain to aid data practitioners choose datasets with more confidence.

\textbf{Ethics.}
Responses were part of a university coursework. Participants consented to the data being used for research when joining the course. No personal data was analysed or reported.

\subsection{Study $2$: Dataset summaries}
\subsubsection{Study $2$ Datasets: the $Set-5$ and $Set-20$ corpora}

We used openly published datasets available as CSV files from three different news sources: \textit{FiveThirtyEight},\footnote{\url{http://fivethirtyeight.com/}} \textit{The Guardian},\footnote{\url{https://www.theguardian.com/}} and \textit{Buzzfeed}.\footnote{\url{https://www.buzzfeed.com/news}} We selected `mainstream' datasets, understandable in terms of topic and content, excluding datasets with very domain-specific language or abbreviations. The datasets had to contain at least $10$ columns and English strings as headers. The datasets varied across several dimensions: value types (strings, integers); topics; geospatial and temporal coverage; formatting of dates; ambiguity of headers, for example abbreviations; blank fields; formatting errors; size; and mentions of personal data.

The sample contained $25$ datasets. We divided them into five groups, each containing: two datasets from \textit{FiveThirtyEight}; two from \textit{The Guardian} and one from \textit{Buzzfeed}. The first of these groups was our first corpus, $Set-$5. $Set-5$ was made of datasets $D1$ to $D5$, which are described in more detail in Table \ref{table:datasets}. We used it in both experiments (see below). The remaining four groups of datasets ($5$ datasets per group, $20$ in total) formed our second corpus, $Set-20$. $Set-20$ consisted of datasets $E1$ to $E20$ and was used only in the crowdsourcing experiment.

Working with $Set-5$ in both experiments allowed us to compare summaries generated by two different participant groups. $Set-20$ enabled us to apply our findings across a greater range of datasets. All datasets are available on GitHub.\footnote{\url{https://github.com/describemydataset/DatasetSummaryData2018}}

\renewcommand{\arraystretch}{1.3}
\begin{table}[H]
\scriptsize
\centering
\begin{tabular}{p{100mm}}
{\small Dataset - Topic - Example characteristics}\\
\hline
\textbf{$D1$ Earthquakes}:
>10k rows, dates inconsistently formatted, ambiguous headers, granular geospatial information\\
\textbf{$D2$ Marvel comic characters}: >16.000 rows, 13 columns, no geospatial information, many string values, limited value rnages, missing values, yearly and monthly values \\
\textbf{$D3$ Police killings}: >450 rows, 32 columns, contains numbers and text, geospatial information (long/lat as well as country, city and exact addresses), personal data, dates as year/month/day in separate columns, headers not all self-explanatory, some domain-specific language   \\
\textbf{$D4$ Refugees}: 192 rows, 17 columns, mostly text values, formatting inconsistencies, ambiguous headers, identifiers, geospatial information (continent/region/country), no temporal information  \\
\textbf{$D5$ Swine flu}: 218 rows, 12 columns, formatting inconsistencies, geospatial information (countries as well as long/lat), links to external sources, identifiers (ISO codes), some headers not straightforward to understandable\\
\end{tabular}
\caption{Datasets in $Set-5$}
\label{table:datasets}
\end{table}

\subsubsection{Study $2$: Lab-based experiment}
\label{sec:methodslab}

The objective of this experiment was to generate summaries of datasets written by data-literate people, who were unfamiliar with the datasets they were describing. Our assumption was that by asking people to summarise datasets unknown to them, they would create summaries that are relatable to a broad range of data users, and would be less biased in their descriptions than people who had been working with that data in the past, or had created it themselves. Each participant was asked to summarise the datasets from $Set-5$ as explained below. Having multiple summaries for the same datasets allowed for more robust conclusions.

\noindent \textbf{Pilot.}
We first conducted a pilot study with one dataset, six participants and different task designs. The aim was to get an understanding of core task parameters, such as the time allocated to complete the task and basic instructions about the length and format of the summaries. These parameters were important to use as we wanted people to report only the most important features of datasets rather than try to document everything they could see. We experimented with several task durations and varying restrictions on the number of words of the summaries. We did not impose any restrictions on the writing (e.g., full text, short notes, bulleted lists etc.). Based on the pilot, we decided to ask participants to write summaries of up to $100$ words with no time limit.

\noindent \textbf{Recruitment.}
We recruited participants who would be the primary target audience for textual data summaries, called \textit{`data practitioners'} for the purpose of this work. While some of the subjects were very experienced in data handling, we chose not restrict participation to formally trained data scientists, as the majority of people working with data are domain experts.\citep{Boukhelifa:2017:DWC:3025453.3025738, DBLP:conf/ercimdl/KernM15}.

Our participants declared to have either received some training in using data or work with data as part of their daily jobs. Previous research has shown that this group are depending on summaries to select datasets with confidence \citep{DBLP:journals/corr/abs-1801-04971,Laura}. By contrast, most data scientists and engineers can easily resort to a range of specialist techniques such as exploratory data analysis to make sense of new datasets.

We recruited participants through a call on social media via the first author's institution. The call was published on the institution's website and a link to the call was posted on Twitter. The Twitter account had at the time of the study over $38.9k$ followers, with $23.025$k impressions, $435$ interactions and $91$ retweets.\footnote{\url{https://support.twitter.com/articles/20171990}} Our sample consisted of n=$30$ participants ($19$ male and $11$ female), all based in the UK at the time of the study. Two thirds of them were UK nationals (n=$20$) and had a Bachelor or Masters level education (n=$26$). All sessions were carried out between July and August $2017$.

\noindent \textbf{Process.}
Respondents to the study call were contacted via email to receive an information sheet. We arranged a time for the experiment at the author's organisation with those who volunteered to take part in the study. The task was formulated as follows: \textit{We ask you to describe the datasets in a way that other people, who cannot see the data, can understand what it is about.}

Participants could open the CSV files with a software of their choice; we suggested MS Excel or Google Sheets. We asked them to describe all five datasets in up to $100$ words, one dataset at a time, in a randomised order, in a text document.

\noindent \textbf{Analysis.}
We collected $150$ summaries, $30$ per dataset. In our analysis we focused on the following aspects: (i) \textit{form} (e.g. full sentences, bullet points etc.) and \textit{length} of the summaries; (ii) \textit{information types} people consider relevant for data sensemaking; and (iii) specific \textit{summary attributes}.

To get a sense of the surface form of the summaries, we counted how many of them used full sentences, bullet points, tables or a mixture of the three. For their length we counted the number of words using the word-count feature in a text editor. To derive information types and summary attributes, two of the authors independently analysed the data inductively using grounded theory to allow themes to emerge \citep{thomas2006general}. We ascribed open codes in an initial data analysis and explored the relationships between the codes in a further iteration (axial coding). We identified higher-level categories by examining properties that were shared across the codes. We adopted this approach because of the open nature of the research questions. We aimed to identify the composition of summaries produced by our participants and understand the relative importance of particular attributes.  We used NVivo, a qualitative data analysis package for coding. In each of the two iterations, we cross-checked the resulting codes, refined them through discussions with two senior researchers, and captured the results in a codebook. We documented each code with a description and two example quotes. Two senior researchers reviewed the conflict-prone codes based on a sample of the data. The unit of analysis was a summary (n=$150$) for the same dataset.

\begin{figure}
\centering
%\begin{footnotesize}
\includegraphics[width=130mm, scale=0.5]{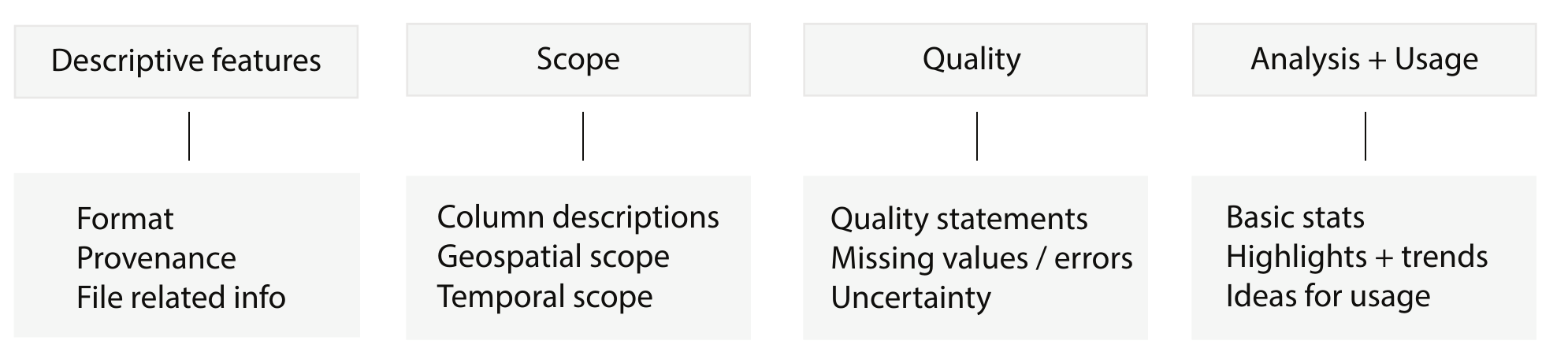}
%\end{footnotesize}
%\vspace{-3mm}
\caption{Information types and emerging attributes from the thematic analysis of the lab summaries}
\label{fig:codingtree}
%\vspace{-3mm}
\end{figure}

\noindent \textbf{Ethics.} The lab experiment was approved by our institution's Ethical Advisory Committee under ERGO Number 28636. Informed written consent was given to the participants prior to the experiment.

\subsubsection{Study $2$: Crowdsourcing experiment}

Following the lab experiment, we undertook a data summaries experiment on the crowdsourcing platform CrowdFlower (now Figure Eight).\footnote{\url{https://www.figure-eight.com/}} We used both dataset corpora, $Set-5$ and $Set-20$ and asked crowd workers to produce summaries of $50$ to $100$ words.

Through the crowdsourcing experiment we were able to reach out to a much larger number of participants to create summaries for more datasets. Using the five datasets from $Set-5$ in both experiments allowed us to compare the characteristics of summaries produced by data practitioners and the crowd.

Existing research suggests crowdsourcing platforms are a feasible alternative to the lab for our purposes. Previous studies have considered related tasks such as text writing \citep{Bernstein:2015:SWP:2808213.2791285}; text summarisation \citep{borromeo2017crowdsourcing,marcu2000theory}; and data analysis \citep{lin2013selecting, DBLP:journals/tvcg/WillettGSHA13}.

\noindent \textbf{Recruitment.} Participants were crowd workers registered on CrowdFlower. We limited the experiment to $Level 2$ crowd workers from native-English speaking countries.\footnote{$Level 2$ workers are workers who have reached a certain level of performance in their previous work.}

\noindent \textbf{Process.} Crowd workers had to describe five datasets (either the datasets $Set-5$ or one of the four groups from $Set-20$) in $50$ to $100$ words. The length of the summaries was informed by the lab experiment. We also included $12$ short qualification questions, assessing basic reading, reasoning and data literacy skills to make sure workers have the capabilities to complete the task.

We used the same basic task description as in the lab: \textit{We ask you to describe the datasets in a way that other people, who cannot see the data, can understand what it is about.}, but included some additional information. Paid microtask crowdsourcing works well when the crowd is provided with a detailed description of the context of the task they are asked to complete. For this reason, we also showed participants step-by-step instructions, tips, a picture of a dataset and examples of corresponding summaries.

Just like in the lab experiment, participants were free to structure their summaries as they saw fit. However, they were shown three examples, presented as text; a list; and a combination of list and text. The minimum time allowed to summarise five datasets was $15$ minutes; the maximum time was $60$ minutes. Both settings were informed by the lab experiment.

The outputs were, for each worker, five textual summaries for five datasets. To minimise spam, we prevented copy-pasting of content and validated a random selection of ten words from each answer against an English language dictionary, requiring a $60\%$ matching threshold to be accepted.

We recruited $30$ crowd workers for $Set-5$ and $20$ crowd workers for $Set-20$ (five workers per each group of five datasets from $Set-20$). Workers were allowed to do only one task i.e. summarise five datasets. They were paid \$$3.00$ per task. From the lab we learned that the task duration is likely to be around $25$ to $35$ minutes, which was confirmed in an early pilot on CrowdFlower.

A screenshot of the CrowdFlower task is included on the GitHub repository created for this work\footnote{\url{https://github.com/describemydataset/DatasetSummaryData2018}}.

\begin{figure*}[h]
\centering
\begin{footnotesize}
\includegraphics[width=140mm]{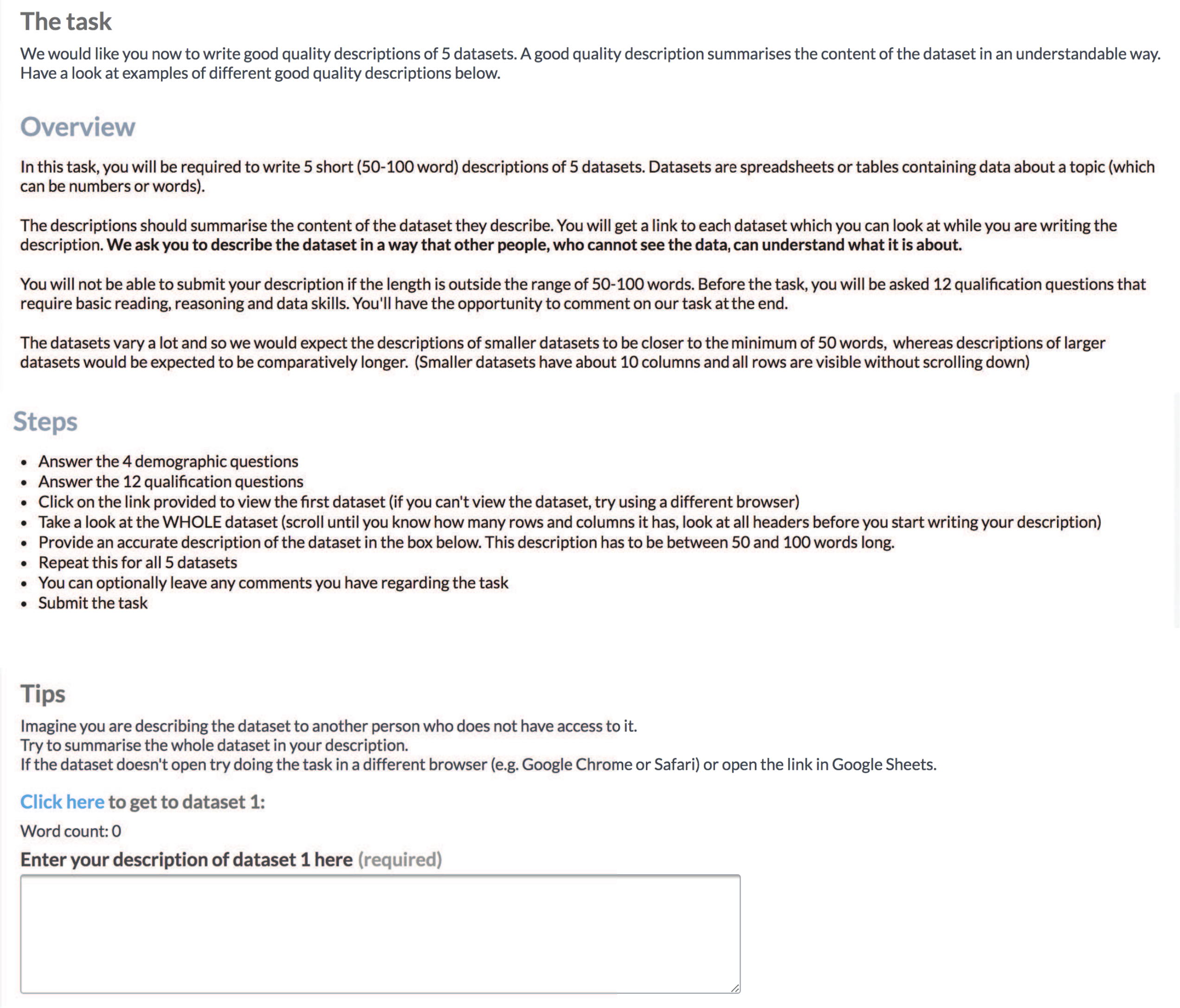}
\end{footnotesize}
%\vspace{-3mm}
\caption{CrowdFlower task instructions in the crowdsourcing experiment}
\label{fig:crowdflower}
%\vspace{-5mm}
\end{figure*}

\noindent \textbf{Analysis.}
We collected a total of $250$ crowdsourced summaries and manually excluded those which were obvious spam or off-topic. This resulted in $120$ summaries for the five datasets in $Set-5$ (on average $24$ summaries per dataset) and $90$ summaries for the $20$ dataset in $Set-20$ (between four and five summaries per dataset). We analysed: (i) the form and length of the summaries; and (ii) the summary attributes, grouped according to the information types identified in the lab experiment. On both accounts we used the same methods as in the lab experiments (see Section \ref{sec:methodslab}). We also looked at differences between the two participant groups for the summaries of $Set-5$ and across datasets for all $25$ datasets from the two corpora.

\noindent \textbf{Ethics.}
This experiment was approved by our institution's Ethical Advisory Committee under ERGO Number 29966. Consent was given by crowd workers previous to carrying out the task.

\section{Findings}
\label{sec:findings}
We report on the results of the two studies. The first study used $269$ data-search diaries by $69$ students to understand what data attributes are relevant in dataset selection, and, hence, how data summaries should look like to be useful to their readers. The second study analysed what attributes people choose to describe previously unknown datasets, based on a total of $360$ data summaries for $25$ datasets created by $80$ participants. We compare both sets of attributes, discuss differences in summary creation across datasets and summary authors, and highlight common themes and characteristics.

\subsection{STUDY $1$: DATA-SEARCH DIARIES}
%authoritative

The analysis of the data-search diaries was performed to \textit{complement} the results of the summary analysis \citep{bryman2006integrating}. In their diaries, the students explicitly answered questions about their thought processes and their rationales when selecting data to work with.

The data attributes emerging from this analyise are listed below, and analysed in more detail in the discussion when we compare the results of our studies with existing metadata standards. We grouped them according to the three high-level themes identified in \cite{Laura}: \textit{relevance}, \textit{usability} and \textit{quality} and describe the topics that emerged within these. Some of the attributes were mentioned by participants in the context of several themes, which emphasises their importance.

\begin{table}[H]
\footnotesize
\centering
\begin{tabular}{p{100mm}}
%{\small}\\
%\hline
\textbf{\underline{Relevance\vphantom{y}}}:
%\vspace{1mm}
%\hline
\begin{itemize}[noitemsep,topsep=5pt]
\item Scope of the data
\begin{itemize}[noitemsep]
%\scriptsize
\item geographical scope
\item temporal scope
\item basic statistics (e.g. counts, value ranges, unique values etc.)
\end{itemize}
\item Granularity (e.g. number of traffic incidents per hour, day, week etc.)
\item Documentation (e.g., understandability of variables, samples)
\item Context
\end{itemize}\\
%\hline
\textbf{\underline{Usability}}:
%\vspace{1mm}
%\hline
\begin{itemize}[noitemsep,topsep=5pt]
\item Format (e.g. CSV, PDF, encodings etc.)
\item Size
\item Documentation
\item Language (e.g. used in headers or for string values)
\item Comparability (e.g., identifiers, units of measurement)
\item References to connected sources
\item Access (e.g. license, API)
\end{itemize}
\\
%\hline
\textbf{\underline{Quality}}:
%\vspace{1mm}
%\hline
\begin{itemize}[noitemsep,topsep=5pt]
\item Provenance (e.g. authoritativeness, context and original purpose)
\item Accuracy (i.e. data is correct)
\item Cleanliness (e.g. well-formatted, no spelling mistakes, error-free etc.)
\item Completeness (e.g. missing values)
\item Timeliness (e.g. how often is it updated)
\item Methodology
\end{itemize}
\end{tabular}
\caption{Selection criteria}
\label{table:datasets}
\end{table}

\begin{comment}
\setlength{\textfloatsep}{0.1cm}
\begin{table} [H]
  \centering
  \scriptsize
  \begin{tabular}{p{1cm}p{5.3cm}p{4mm}}
    \textsl\small{\textsc{Theme}}
   & \textsl\small{\textsc{Attribute}}
   &  \textsl\small{\textsc{\%}} \\
    \midrule
    Relevance  & coverage (topical, geographical, temporal)\newline granularity\newline comparability to other datasets \newline  understandability of variables\newline  sample \newline context and purpose of the data & 36 \newline 19 \newline 14 \newline 14\newline 8  \newline 6 \\
       \midrule
    Usability  & format (data type, structure) \newline understandability, documentation \newline access (license, API's) \newline comparability (identifiers) \newline  size \newline units of measurement\newline  temporal (up-to-date, time of publishing) \newline language &  44 \newline 17 \newline 11 \newline 11  \newline 7 \newline 6\newline 4 \newline 2 \\
       \midrule
    Quality  & provenance \newline accuracy
    \newline temporal
    \newline comprehensiveness / completeness \newline cleanliness \newline documentation  \newline methodology
\newline references to connected sources\newline comparability to other datasets  \newline size & 28 \newline 14 \newline 13 \newline 12 \newline 9   \newline 9  \newline 8 \newline 5   \newline 3 \newline 3 \\
  \end{tabular}
  \caption{Themes and attributes used to select datasets}~\label{tab:selectioncriteria}
\end{table}
\end{comment}

\subsubsection{Relevance}
Two of the most prevalent attributes were the \textit{scope of the data} (in terms of what it contains) and its \textit{granularity}. They were mentioned in $36\%$ and $19\%$ of responses, respectively. For example, the scope sometimes referred to the geographical area covered by the dataset, while the granularity described the level of detail of the information (e.g. street level, city level, etc.) Some participants mentioned \textit{basic statistics} such as counts, averages and value ranges as a useful instrument to assess scope.

Interestingly, $14\%$ of the diaries noted the relative nature of relevance (echoing discussions in the literature \citep{mizzaro1997relevance}) and the need to consider multiple datasets at the same time to determine it. To a certain extent, this could be due to the nature of the task - students were free to choose the topic of the datasets and hence might have had a broader notion of relevance, which allowed them to achieve their goals by interchanging one dataset for another or through a combination of datasets. However, the relation to other sources was mentioned in other categories as well, which reinforces the need for tools that make it easy for data users to explore more than one dataset in the same time and to make comparative judgements. This is also in line with experience reports about data science projects in organisations - making complex decisions often involves working with several datasets \cite{Laura,erete2016storytelling}.
Further attributes from the diaries suggest that a thorough assessment of relevance needs to include easily understandable variables, data samples for fast exploration, as well as insight into the context and purpose of the data.

\subsubsection{Usability}
To determine how usable a dataset is for their task participants mentioned a range of practical issues which, if all available in the desired way, would make working with a dataset frictionless: \textit{format}, \textit{size}, the \textit{language} used in the headers or for text values, \textit{units of measurement} and so on.

\textit{Format} was the most prevalent attribute ($44$\%), though \textit{documentation} and the ability to understand the variables were perceived to impact usability as well (both at $11\%$).

The \textit{size} of the dataset was mentioned primarily in the context of usability rather than basic statistics in relevance. This is probably due to the fact that students were mindfull of the additional effort required to process large datasets.

The participants understood the importance of being able to integrate with other sources, for example through identifiers - $11\%$ of the diaries mentioned this aspect explicitly. In their coursework, the students were asked to use at least two datasets and hence valued data integration highly. In the same time, using multiple datasets is not uncommon in most professional roles \citep{Laura,Convertino:2017:SDP:3027063.3053359}. \textit{Access} to the data was also mentioned in reference to APIs or licences though only around $6\%$ of the times. This low value is a function of our study - students were not looking to source data to solve a fixed problem. Their search for data, documented in the diaries, happened while they were deciding on the topic of their project. If they could not find data for one purpose, they could adjust the project scope rather than having to tackle licensing or access fees.

\subsubsection{Quality}
Participants mentioned unique attributes such as \textit{provenance} - in a broad sense of the term - that would allow judgements around the authoritativeness of the publisher and the original context of the data. At $28\%$ this attribute was ranked much higher than other quality dimensions such as \textit{accuracy}, \textit{completeness}, \textit{timeliness} and \textit{cleanliness}, which are in the focus of many quality repair approaches \citep{DBLP:journals/cacm/WandW96}. The importance of provenance resonates with previous work in data quality \citep{DBLP:journals/jdiq/CeolinGMFHN16, DBLP:conf/amcis/MalaverriMM13}; there is also a large body of literature proposing frameworks and tools to capture and use provenance, though their use in practice is not widespread, for example \citep{DBLP:conf/ipaw/Stamatogiannakis14,DBLP:journals/jwsr/SimmhanPG08}.

Some participants reported to be interested in details of the \textit{methodology} to create and clean the data, including aspects such as the control group, whether a study had been done using randomised trials, confidence intervals, sample size and composition etc. This is in line with an earlier study of ours \citep{Laura}, which pointed out that awareness of methodological choices plays an important role in judging the quality of a dataset with confidence.

In the discussion in Section \ref{sec:discussion} we relate these different data selection attributes to the attributes extracted from the summaries, and compare them to existing guidelines for data publishing and sharing. We identify overlaps between the information needs of people searching for data, who are potential consumers of data summaries, and the information people choose when summarising an unfamiliar dataset.

\subsection{STUDY $2$: DATASET SUMMARIES}
We report on the main findings from the lab and crowdsourcing experiments, covering the three areas mentioned in the methodology in Section \ref{sec:methodology}: (i) summary \textit{form} and \textit{length}; (ii) \textit{information types}; and (iii) detailed \textit{summary attributes}.

\noindent \textbf{Form and length.}
In the lab experiment participants were not given concrete suggestions or examples for surface representation, yet most resulting summaries were presented as text using full English sentences ($64\%$). However, some ($17.3\%$) were structured as a list or presented as a combination of text and lists ($18.6\%$). In the crowdsourcing experiment participants were provided with examples of summaries using these three representations. Their summaries were structured as follows: ($79\%$) used text, a few ($7\%$) were structured as a list, and some ($14\%$) presented a combination of the two.

The average lab summary was $98$ words long (median of $103$). By comparison, the crowd needed on average $63$ words for the same datasets in $Set-5$. It would be interesting to explore how the length of the summaries impacts on their perceived usefulness by readers or on their potential information gain \citep{DBLP:conf/sigir/MaxwellAM17}, also in the context of the summary template we propose in Section \ref{sec:discussion}.

\subsubsection{Information types}
\label{sec:resultsinformationtypes}

We identified four high-level types of information in the lab summaries, which we subsequently used to analyse summary attributes for all $360$ summaries created in the two experiments.

\begin{description}
\item [1. Descriptive attributes] e.g. format, counts, sorting, structure, file-related information and personal data: %

\begin{quote}
%\vspace{-5px}
%(P2)It lists number of refugees, or people in refugee-like situations, identifying the total number of population of concern.\\
(P1) The dataset has 468 rows (each representing one person who has been killed) and 34 columns.\\
(P7) No free text entries and character entries have a structured format.
It contains no personal data\\
(P8) The header pageID appears to be a unique identifier\\
(P21) CSV in UTF encoding. Header and 110172 data rows. 10 columns.\\
%\vspace{-5px}
\end{quote}

\item [2. Scope of the data] which refers to the actual content of the dataset, through column descriptions such as headers or groupings of headers, or references to the geographic and temporal scope:

\begin{quote}
%\vspace{-5px}
(P1) For example, this includes details on the share of ethnicities, in each city, the poverty rate, the average county income etc.\\
(P14)Figures include number of confirmed deaths and proportion of cases per million people.\\
(P2) Some columns have no particular meaning to a non-expert, e.g. columns named "pop", "pov", "country-bucket", "nat-bucket".\\
(P12) Each instance has specific details on the time, geographic location, earthquake's magnitude.\\
%\vspace{-5px}
\end{quote}

\item[3. Quality] which included dimensions such as errors, completeness, missing values and assumptions about accuracy, but also expressions of uncertainty and critique:

\begin{quote}
%\vspace{-5px}
(P1) The precision of the description varies wildly\\
(P14) A link (in some cases two) to the source of the data is provided for each country.\\
(P7) It has column headers all in caps (apart from 'pageID'), which are mostly self-explanatory\\
(P8) Combination of personal data about person killed and demographic data, unclear if this is for area of killing.\\
(P17) The data seems to be consistent and there aren't any empty cells.\\
%\vspace{-5px}
\end{quote}

\item [4. Analysis or ideas for analysis and usage] such as simple data analysis, basic statistics, highlights of particular values or trends within the data:

\begin{quote}
%\vspace{-5px}
(P5) The data does provide the method of how each individual has been killed which can provide an argument for police not using firearms in the line of duty.\\
(P7) There is a significant amount of missing data in the "state" column, but this information should be possible to infer from the "longitude" and "latitude" columns\\
(P8) The file has 13 fields and 16373 records.\\
(P18) The dataset shows that the greatest number of refugees originate from the Syrian Arab Republic. \\
(P30) Killings took place all around America. The people who were killed mostly carried firearms\\
%\vspace{-5px}
\end{quote}

\end{description}

Table \ref{tab:infotypepercentages} shows the percentage of summaries that contained each information type, split by dataset. The four types are not meant to define an exhaustive list - we consider them merely a reflection of the $150$ lab summaries analysed and in Section \ref{sec:limitations} we discuss this limitation of the study. The types are also not exclusive - more than half of the summaries included all four types of information.  \textit{Analysis and usage} was the least frequent information type overall, though some attributes in this category were more popular than others. For example, as we will note later in this section, \textit{basic statistics} were mentioned more frequently in the crowd-generated summaries than in the lab, while trends and ideas for further use were rather low overall, with the exception of some $Set-20$ datasets described by the crowd. We believe the main reason for this is the design of the task. In the lab experiment, the task description might have implied a focus on the raw data and on surface characteristics that could be observed through a quick exploration of the data rather than an extensive analysis. Crowdsourcing requires a higher level of detail in instructions, which included examples of summaries with, among other things, basic statistics.

We present all individual attributes associated with each of the four information types in the remainder of this section.

\begin{table}[H]
\centering
\footnotesize
\pgfplotstabletypeset[color cells]{
x, Total , $D1$ , $D2$ , $D3$ , $D4$ , $D5$
Descriptive attributes,81,67,87,80,83,90
Scope of the data,99,97,100,100,100,100
Data quality,79,80,67,90,77,80
Analysis and usage,64,57,63,73,67,60
}
\caption{Percentages of information types per dataset in $Set-5$, based on $150$ lab summaries}
\label{tab:infotypepercentages}
\end{table}

\subsubsection{Summary attributes}
\label{sec:resultssummaryattributes}
Across the $360$ summaries created in the two experiments we have identified the following attributes:

\renewcommand{\arraystretch}{1.3}
\begin{table}[H]
\scriptsize
\centering
\begin{tabular}{p{78mm}}
{\small \underline{Summary attributes}}\\
%\hline
\textbf{Format}: File format, data type, information about the structure of the dataset \\
\textbf{Provenance}: Where the dataset comes from, such as publisher, publishing institution, publishing date, last update  \\
\textbf{Subtitle}: A high-level one-phrase summary describing the topic of the dataset\\
\textbf{Headers}: Explicit references to dataset headers\\
\textbf{Groupings}: Selection, groupings or abstraction of the headers into meaningful categories, key columns\\
\textbf{Geographical}: Geospatial scope of the data at different levels of granularity\\
\textbf{Temporal}: Temporal scope of the data at different levels of granularity\\
\textbf{Quality}: Data quality dimensions such as inconsistencies in formatting, completeness etc.\\
\textbf{Uncertainty} For example ambiguous or unintelligible headers or values, or unclear provenance\\
\textbf{Basic statistics}:  For example, counts of headers and rows, size of the dataset, possible value ranges or data types in a column\\
\textbf{Patterns/Trends}: Simple analyses to identify highlights, trends, patterns etc.\\
\textbf{Usage}: Suggestions or ideas of what the dataset could be used for\\
\end{tabular}
\caption{Most frequent summary attributes, based on $360$ summaries of datasets from $Set-5$ and $Set-20$}
\label{table:attributes}
\end{table}

%\noindent \textbf{Overview}
Across the two experiments, a summary was commonly structured as follows: (i) a high-level \textit{subtitle} describing the topic of the dataset; (ii) references to dataset \textit{headers} (either the names of the headers or an abstraction of the headers such as a meaningful \textit{grouping}); (iii) a \textit{count} or other descriptive attribute such as possible values in a column; and (iv) \textit{geographic} and \textit{temporal scope}. Amongst other popular attributes were: \textit{quality statements}; \textit{provenance}; and, less frequently, ways to \textit{analyse} or \textit{use} the data.

Here is a summary that exemplifies this:

\begin{quote}
%\vspace{-5px}
(P6) A list of people killed by US police forces in 2015. Data included is location of incident, police department, state, cause of death and whether the victim was wielding a weapon. Detailed and specific data with 34 columns. Useful for drawing parallels between criminal profiling and locations.
%\vspace{-5px}
\end{quote}

Some summaries described the data by talking about the header row as an example:

\begin{quote}
%\vspace{-5px}
(P16) Each row describes one of those `earthquakes': lat, lon, magnitude and location name.
%\vspace{-5px}
\end{quote}

Percentages of these attributes over all summaries can be seen in Table \ref{tab:overviewattributes}, split by experiment and dataset corpus ($Set-5$ and $Set-20$). Attributes such as \textit{subtitle}, \textit{geographical} and \textit{temporal} scope and \textit{headers} were present in a majority of summaries. \textit{Format} was mentioned in more than half of the the $Set-5$ summaries and in $27\%$ of the $Set-20$ summaries. \textit{Basic statistics} were mentioned fairly often as well, in more than half of the $Set-5$ summaries and in $48\%$ of the $Set-20$ summaries.

\begin{table} [h]
\centering
%\footnotesize
\scriptsize
\pgfplotstabletypeset[color cells]{
x,L $Set-5$ ($150$), C $Set-5$ ($120$), C $Set-20$ ($95$)
\raggedright Format,61,52,27
%Format,71,83,47
Provenance,45,24,23
Subtitle,89,86,95
Headers,70,82,80
Groupings,49,70,69
Geographical,73,71,60
Temporal,58,55,56
Quality,55,23,21
Uncertainty,69,16,8
Basic statistics,56,74,48
Patterns/Trends,27,25,52
Usage,15,5,1
}
\caption{Percentages of summaries created in the lab ($L$) and via crowdsourcing ($C$) that mention summary attributes. Darker fields have higher percentages. Numbers in brackets refer to the number of summaries analysed in each category.}
\label{tab:overviewattributes}
\end{table}

Table \ref{tab:prevalenceexperts} elaborates on the distribution of summary attributes in the lab summaries ($150$ summaries in total, $30$ per dataset). Across the five datasets analysed, \textit{subtitle}, \textit{format} and \textit{headers} were mentioned consistently in more than $55\%$ of the cases. \textit{Basic statistics} and \textit{quality} achieve slightly lower scores ($47\%$ and higher). We discuss differences in scores between datasets as well as the attributes that showed greater variation later in this section.

\begin{table} [h]
%LONG:
\centering
\footnotesize
\pgfplotstabletypeset[color cells]{
x, $D1$ , $D2$ , $D3$ , $D4$ , $D5$
\raggedright Format,60,63,57,60,67
%Format,67,73,70,67,77
Provenance,23,47,10,53,90
Subtitle,83,87,87,83,90
Headers,60,87,80,63,60
Groupings,13,40,70,53,30
Geographical,90,0,90,90,93
Temporal,87,37,87,33,47
Quality,57,50,47,60,60
Uncertainty,73,60,80,67,67
Basic Stats,53,73,63,47,43
Patterns/Trends,23,20,33,27,30
Usage,17,20,17,7,17
}
\caption{Percentage of lab summaries containing respective attributes, per dataset from $Set-5$. Darker fields have higher percentages}
\label{tab:prevalenceexperts}
\end{table}

Table \ref{tab:prevalencecrowd5} illustrates the distribution of attributes in the $120$ summaries by the crowd for the datasets in $Set-5$ ($24$ summaries per dataset on average). The most prevalent attributes are slightly different than the ones observed in the lab: \textit{subtitle}, \textit{format} and \textit{headers} remain important, but \textit{basic statistics} are more consistently mentioned than in the other experiment. In the same time, the crowd focused on \textit{groupings} of headers as well, much more so the data practitioners who participated in the lab experiment - overall $70\%$ of the crowd-generated summaries of $Set-5$ mentioned this attributes, compared to only $49\%$ of the lab summaries (see Table \ref{tab:overviewattributes}); the scores for the individual datasets were between $62\%$ and $78\%$ on CrowdFlower and $13\%$ and $70\%$ in the lab.

The $Set-20$ summaries created by the crowd reinforce some of these trends (see Table \ref{tab:overviewattributes}). When looking at the $20$ datasets from $Set-20$, \textit{subtitle}, \textit{geographical} and \textit{temporal} scope and \textit{headers} are mentioned in the majority of summaries, just like in the summaries of $Set-5$. \textit{Groupings} seem to be popular among crowd workers across all datasets  - $69\%$ of the $95$ $Set-20$ summaries mention them, compared to $70\%$ of the $120$ $Set-5$ summaries, but only $49\%$ of the lab summaries. By contrast, $Set-20$ summaries showed greater variation in: \textit{format} ($27\%$ vs $52\%$ of the crowd-produced $Set-5$ summaries vs $61\%$ for the lab-produced $Set-5$ summaries) and basic statistics ($48\%$ vs $74\%$ of the crowd-produced $Set-5$ summaries vs $56\%$ for the lab-produced $Set-5$ summaries). Aside the popularity of \textit{groupings}, a second surprising result was the popularity of \textit{patterns/trends} aspects - this attribute was mentioned in only $27\%$ of the lab summaries and $25\%$ of the crowd summaries for the same $Set-5$ datasets (see Table \ref{tab:overviewattributes}, but in $52\%$ of the $Set-20$ summaries. This goes against our basic assumption that the task instructions suggested a focus on raw data and surface characteristics. Later in this section, we will unpick the summaries that referred to \textit{patterns/trends} aspects to understand how this difference came about.

%\TODO{Discuss high percentage of analysis attributes in set-20 and L: done later on, in "differences between set 5 and set 20 paragraph - enough?}

\begin{table} [h]
\centering
\footnotesize
\pgfplotstabletypeset[color cells]{
x, $D1$ , $D2$ , $D3$ , $D4$ , $D5$
\raggedright Format,55,50,57,48,52
%Format,90,73,91,83,76
Provenance,20,23,0,17,62
Subtitle,75,95,91,70,100
Headers,75,86,83,78,86
Groupings,65,73,78,70,62
Geographical,95,0,91,78,90
Temporal,90,68,61,0,62
Quality,45,23,22,17,10
Uncertainty,35,9,17,4,14
Basic Stats,75,77,83,78,57
Patterns/Trends,30,32,17,26,14
Usage,0,9,4,4,5
}
\caption{Percentage of crowdsourced summaries from $Set-5$ containing respective attributes, per dataset. Darker fields have higher percentages}
\label{tab:prevalencecrowd5}
\end{table}

%\begin{table*} [h]
%\centering
%\tiny
%\pgfplotstabletypeset[color cells]{%
% x, $E1$, $E2$, $E3$, $E4$, $E5$, $E6$, $E7$, $E8$, $E9$, $E10$, $E11$, $E12$, $E13$, $E14$, $E15$, $E16$, $E17$, $E18$, $E19$, $E20$
% \raggedright Subtitle,60,100,100,100,100,100,100,100,80,100,100,100,100,100	,100,75,100,100,75,100
% Format,0,80,50,25,20,60,75,50,40,60,25,50,20,60,75,25,40,80,75,25
% Provenance,20,20,0,0,40,60,25,0,0,0,0,25,80,20,50,25,0,0,75,0
% Geographical,100,100,100,100,40,80,25,50,80,20,0,75,100,100,25,50,20,20,0,75
% Temporal,60,80,25,50,20,100,50,25,60,100,100,100,20,0,25,50,40,40,50,75
% Headers,100,100,100,25,80,80,75,100,80,100,50,75,100,100,75,100,80,40,50,75
% Quality,0,0,25,50,0,20,0,50,20,20,0,25,40,0,0,25,0,40,50,25
%Uncertainty,
%Groupings,
% Basic Stats,20,80,50,25,60,60,75,25,20,60,50,0,20,60,75,25,20,80,100,25
% Analysis,40,80,50,50,40,40,50,50,20,60,75,50,40,60,50,75,40,60,50,50
% Usage,0,0,0,0,0,0,0,0,0,0,0,0,0,0,0,25,0,0,0,0
%}

%\caption{CUT:Percentage of crowdsourced summaries from $Set-20$ containing respective attributes, per dataset. Darker fields have higher percentages}
%\label{tab:set20}
%\end{table*}

\noindent \textbf{Differences between lab and crowd summaries}\\
%\subsubsection{Differences between lab and crowd summaries}\\
The five datasets from $Set-5$ were used in both experiments. As noted earlier, for the lab experiment, our sample consisted of $150$ summaries from $30$ participants, while for the crowdsourcing experiment we used a sample of $120$ summaries from $30$ participants (in the crowdsourcing experiment we had to eliminate some summaries due to spam).

We compare the distributions of attributes in the two experiments shown in Table \ref{tab:overviewattributes}. For the five datasets in $Set-5$, \textit{provenance} appears more frequently in the summaries created in the lab ($45\%$ vs $24\%$). We believe this to be due to the fact that participants were more data savvy and so placed a greater importance on where a dataset originates from. A similar trend was observed in the data-search diary, where participants were MSc students reading data science, computer science or statistics. We assume the same applies for \textit{quality} statements ($55\%$ vs $23\%$) and ideas for \textit{usage} ($15\%$ vs $5\%$), whose appreciation may equally require a certain level of experience with data work which was not given in the crowdsourcing setting. In the same time, the crowd appreciated attributes such as \textit{groupings} of headers ($21$ point difference) and \textit{basic statistics} ($18$ points) more. This demonstrates that the crowd had a fair level of data literacy and does not focus only on features that can be easily observed such as \textit{subtitle}, \textit{format} and \textit{headers}. As noted earlier, when looking at summaries for $20$ other datasets, \textit{groupings} remained popular, but \textit{basic statistics} dropped to a lower level than in the lab ($48\%$). We believe this calls for additional research to understand the relationship between the capabilities of summary authors and the aspects they consider important in describing datasets to others.

Looking at the distribution of summary attributes over $Set-5$ (Table \ref{tab:prevalenceexperts}), geospatial attributes, as well as provenance appear to have the highest dependency on the dataset. $D5$ differed from the other four datasets in the corpus by including an entire column titled \textit{'Sources'}, displaying links to the source from which the values were taken from - this is likely the reason why $90\%$ of the $30$ data practitioners and $17$ crowd workers mentioned it in their summaries. $D2$ similarly included a header called \textit{'Page id'} pointing to the source of the data - this was less easy to spot by the crowd workers, who talked about \textit{provenance} only $17\%$ of the time.

We believe that geospatial attributes might in reality be more consistent for most datasets - four out of five datasets achieved consistently high scores in this category. $D2$ was set in a fictional universe and may have therefore not prompted participants to discard any geospatial considerations.

\noindent \textbf{Differences between $Set-5$ and $Set-20$ summaries}\\
The crowdsourcing experiment used two corpora: $Set-5$ with the same five datasets used in the lab and $Set-20$ with $20$ datasets. The reason to include a second corpus, albeit with fewer summaries per dataset ($95$ summaries in total, four to five summaries per dataset) was to explore how the main themes that emerged from the $270$ summaries of $Set-5$ generalise across datasets.

Compared to the $Set-5$ crowd-generated summaries, $Set-20$ shows a higher prevalence of \textit{subtitles} ($95\%$ vs $86\%$) and \textit{patterns/trends} ($52\%$ vs $25\%$) and lower scores for \textit{format}, \textit{geographical} scope and \textit{basic statistics} (see Table \ref{tab:overviewattributes}).

We looked at each of the $20$ datasets from Set-$20$ to understand where these differences might come from. $Set-20$ contained a higher number of datasets with clearly identifiable \textit{subtitles}, which explains the higher score. The datasets overall had fewer attributes representing \textit{format} and \textit{basic statistics}. Many $Set-20$ datasets either did not contain any \textit{geographical} information or were clearly associated with a country or region that is not mentioned explicitly -  for instance, $E10$ is about the UK's House of Commons, but there are no geospatial values in the dataset. The popularity of \textit{patterns/trends} in $Set-20$ points to another dependency of summary content on the dataset - both in the lab and on CrowdFlower, the summaries of the $Set-5$ datasets were consistent along this dimension.
For instance, E11 explicitly mentions statistical content such as 'the median' as a header,
other summaries with a high percentage of patterns/trends attributes tend to display clear trends or rankings and therefore afford quick judgements, for instance "the country with the highest human development index". The same counts for datapoints that stand out that get highlighted in a summary. For example in the example of a dataset (E10) that contains salaries and expense claims from members of the British Parliament House of Commons which shows claims for a lawn mower, amongst other claims.

Just like the other summaries produced by crowd workers, \textit{usage}, \textit{provenance} and \textit{quality} were not mentioned very often, which we believe is due to the level of data literacy in the experiment. In addition, we noted that $Set-20$ provenance was often not recorded when the context or origin of the dataset was very opaque - for example, $E4$ had mainly numerical values describing the elderly population worldwide - or in connection to uncertainty about the provenance - e.g. $E12$ was about US weather data, but did not make any reference to the source of the data.

\subsubsection{Summary attributes in detail}
In the previous section we presented a series of high-level findings across the two experiments and differences across datasets and participant groups. In this section, we discuss summary attributes individually and give additional details and example summaries.

%\textit{Descriptive attributes.}\\
\noindent \textbf{Format and file related information.}\\
\textit{Format}. The file format and references to the structure of the dataset were explicitly mentioned in more than $60\%$ of all lab summaries and in about half of all $Set-5$ crowdsourced summaries. The mentions of file format or data type drop for $Set-20$ to $27\%$.

\textit{File related information}. The summaries contained other attributes that described the file beyond its actual content, which refers to descriptive attributes as mentioned in the overview section on information types represented in the summaries.
That included attributes such as: the type of values in a column;
statements about the size of the file; mentions of licence ($3\%$ of the lab summaries and none of the crowdsourcing summaries); sorting of values; redundancies in the data; formatting; and unique identifiers.
The valuetype of a column was mentioned in $18\%$ of all lab summaries, and in $23\%$ for $Set-5$, $15\%$ for $Set-20$.

There were also mentions of personal data in this category, as they describe a characteristic of the data rather than the data itself. Personal data was mentioned by 20\% of the participants and mostly mentioned in connection to D3 which contained names of people in a police crime. We assume this is due to the fact that in the context of our task and the type of data we used (aside from D3), personal data was not an category that our participants were prompted to think of.

\noindent \textbf{High-level subtitle.}
Close to $90\%$ of all summaries started with a high-level \textit{subtitle} which gave the reader a quick first impression of what the dataset was about. In some cases \textit{subtitle} referred to a key column $6-7\%$ or, more often, to the geospatial scope ($48\%$ of the lab summaries and ~$35\%$ of the crowdsourced summaries), or to the temporal scope of the dataset ($33\%$ of the lab summaries, $17\%$ $Set-5$ crowdsourced summaries and $19\%$ of $Set-20$).

%\vspace{3px}

\begin{quote}
(P1) This dataset, in csv format, describes police killings in what appears to be the USA in 2015. %Also mentions confidence/certainty (i.e. "appears")
\end{quote}

\begin{quote}
%\vspace{-3px}
(P11) Dataset of characteristics of Marvel comic book characters from the earliest published comics to around 2013.
%\vspace{-3px}
\end{quote}
\begin{quote}
%\vspace{-3px}
(P13) This dataset describes the time, geographical location and magnitude of earthquakes in the United States.
%\vspace{-3px}
\end{quote}

\noindent \textbf{Column descriptions.}
%\noindent \textbf{Headers and groupings of headers}
A majority of summaries explicitly mentioned the headers of the dataset (70\%). This was consistent through all summaries done by the same participant -- which points to the fact that this feature is not dependent on the underlying dataset. About half of all summaries show some type of grouping or abstraction of the headers. Participants typically mention a selection of headers
and group them according to meaningful categories, as can be seen below:

\begin{quote}
%\vspace{-3px}
(P14) 34 variables, which comprehend personal information about the victim, place (inc. police department) of the incident, details about the incident, socio-demographic of the place.
(P15) Fields: Demographic data (name, age, gender, race), date (month, year, day), incident details (cause of death, individual armed status - categorical), county details (population, ethnicity), law enforcement agency, general reference data.
%\vspace{-3px}
\end{quote}

Similarly, a common strategy is the identification of a key column, which is the focus of the dataset:

\begin{quote}
%\vspace{-3px}
(P11) For the victims, the metadata records their age, gender, ethnicity, address. The place and time of their death, as well as the cause of death and police force responsible are also recorded.
(P23) We are given useful information about each earthquake, specifically: latitude, longitude of the event, magnitude of the earthquake, a unique identifier for each earthquake called `id', when the data was last updated, the general area the earthquake took place, the type of event it was, the geometrical data and if it took place in the US we are given the state it occurred in.
%\vspace{-3px}
\end{quote}

Some participants use the actual header name, others use a more descriptive version of the header. Many list the headers, together with qualifying information about them and/or possible values and ranges in a column.

\begin{quote}
%\vspace{-3px}
(P30) It lists more than 15000 characters with their fictitious name and the real name in the comic. The data set records whether they are alive or dead characters, their gender, their characteristics (like: hair and eye colour). The data set records if the character has a secret identity [..] (and) whether the particular character has a negative or positive role.
%\vspace{-3px}
\end{quote}

\noindent \textbf{Geographical information.}
\textit{Geospatial} aspects were very common in summaries across datasets and participant groups. In $Set-5$, the exception was $D2$, which described characters in a fictional world. They referred to different types of locations, including provenance (where the data comes from), coverage of the data itself (e.g. data from a particular region), and format, at varying levels of granularity.  Summary authors often used higher-level descriptions of the relevant values, for example \textit{"for most countries in the world"} or \textit{"across the world" } to describe key columns with a wide range of country names.

\begin{quote}
%\vspace{-5px}
(P17) The data goes down to country and includes country codes, the area and region.
(P1) location (provided by latitude and longitude measurements)
(P2) location (in latitude and longitude, but also in descriptive text about location relative to a city)\\
(P7) Each observation refers to a unique country, using country codes
%\vspace{-5px}
\end{quote}

\noindent \textbf{Temporal information.}
\textit{Temporal} aspects were mentioned in connection to: time mentioned in the data, the publishing date, the last update and the time the data was collected, all at different levels of granularity. The numbers reported as here include only temporal attributes that refer to the temporal scope of the data itself and not to publishing date or last updates which were included in \textit{provenance}. %\TODO{makes sense?}

Often summaries refer to both \textit{"date"} and \textit{"time"}, meaning the time of the day and the day that a particular event in the data occurred.

We found differences depending on the datasets: in the $Set-5$ lab summaries, for example, time was most often mentioned in relation to $D1$ and $D3$ ($87\%$) and less often in connection to $D2$ and $D4$ ($<40\%$). $D3$ had three date columns separating day, month and year from each other which might prompt including this information in the summaries. $D1$ had high inconsistency in formatting dates and included two types of temporal information: when the earthquake took place and when the specific row was updated. $D1$ displayed a relatively high overlap between time and uncertainty ($30\%$ of all mentions of time were connected to uncertainty).%, while D3 showed the least overlap with $7\%$.
This points to inconsistencies in formatting of dates in D1 and to potentially confusing headers called "time" and "updated", which show a mixture of dates and times. We assume this contributes to the varying prevalence of time in the summaries, which can be seen in Table~\ref{tab:prevalenceexperts}.
%Summaries for $D2$ and $D4$ displayed significantly lower percentages of temporal information. %marvel and refugees
$D4$ on the other hand did not contain temporal information explicitly which explains the significantly lower percentage. This was reflected in the crowdsourced summaries for $D4$. $D2$ did contain temporal information (year and month), however it describes fictional comic characters which may lead to placing less importance on the temporal information represented in the data. \\

%\TODO{new}
\textit{Temporal provenance.}
We further saw mentions of updates of the data, which we define as temporal provenance. This was present in $20\%$ of all lab summaries and in $6\%$ of the $Set-5$ and $12\%$ for $Set-20$ crowdsourced summaries. It describes mentions of time that can be used to determine the relevance or quality of the data, such as:
\begin{quote}
%\vspace{-3px}
(P30) The data set for confirmed cases of flu was last updated on 20/01/2010.\\
(P1) It is unclear whether this data is up to date, as there are no details on when this is from.
%\vspace{-3px}
\end{quote}

\noindent \textbf{Quality statements and uncertainty.}
Statements about uncertainty and quality were common in 70\% of the lab summaries.
Among the most popular words in this category were `unclear' and `missing'.
The emerging themes connected to quality were features such as inconsistencies in formatting (e.g. dates), completeness, as well as statements about missing understandability (such as ambiguous or unintelligible headers or cells), as well as unclear provenance and authoritativeness of the source.

We further grouped uncertainty statements into six categories related to: completeness, precision, definitions, relations between columns, temporal and geospatial attributes, and methodology.

\textit{Completeness} included statements about the representativeness, comprehensiveness and scope of the data, in addition to general statements about missing values:
\begin{quote}
%\vspace{-10px}
(P4) Unclear how representative this list is of total population/whether this list is total population
%\vspace{-5px}
\end{quote}

\begin{quote}
%\vspace{-5px}
(P13) The dataset appears to be missing data from some of the countries.
%\vspace{-5px}
\end{quote}

\textit{Accuracy} referred to inconsistencies in the data, for instance in units of measurements, or variations in the granularity of cell values.

\begin{quote}
%\vspace{-5px}
(P13) The precision of the description varies wildly (eg. 23 km NE of Trona versus Costa Rica).
%\vspace{-5px}
\end{quote}

\textit{Definitions} were a common theme within uncertainty, such as unclear meaning of headers or identifiers, acronyms or abbreviations or other naming conventions. This seemed especially important for numerical values as there is often no further context given to a cell value or no information provided on what missing values mean:

\begin{quote}
%\vspace{-5px}
(P24) Uncertainty what missing values mean was noted:
This dataset is clear and is very dense although it is possible that the zero values in the set denote that the data could not be obtained.
%\vspace{-5px}
\end{quote}

\begin{quote}
%\vspace{-5px}
(P27) It's not clear how the 'magnitude' is measured, presumably it's the Richter scale but that isn't specified.
%\vspace{-5px}
\end{quote}

\textit{Relations} between columns, or dependencies between columns were mentioned within uncertainty.\\

\begin{quote}
%\vspace{-5px}
(P1) It is unclear whether these are civilians who have been killed by police, or policemen who have been killed by, though I assume it is the former.
%\vspace{-5px}
\end{quote}

\textit{Temporal and geospatial attributes} within uncertainty referred to unclear levels of aggregation or granularity of these attributes and potential ranges of values within a column. Furthermore, it seemed to be often unclear whether the data was up-to-date, and whether events in the data represent the time these were recorded or the time these happened. $19\%$ of all mentions of uncertainty are connected to time and $28\%$ to location:

\begin{quote}
(P14) All the data is related to 2015, although I do not know whether all the data about this year is contained in this dataset.

(P1) It is unclear to me whether these details are from the city, county, or state level.
\end{quote}

\textit{Methodology}:
Uncertainty statements also presented questions related to methodology of data collection and creation. These covered aspects such as: \textit{how were these numbers calculated, are they rounded, how was the data collected, what was the purpose of the data?}
Some of these aspects refer to the provenance of the data and the importance of awareness of methodological choices during data creation was also found to be an explicit selection criteria in the results of the diary study.

\noindent \textbf{Basic statistics.}
%\textit{Basic stats:}
%relations to the datasets
Basic statistics about the dataset were one of the most prevalent features in the \textit{analysis and usage} category (mentioned by 77\% of all participants, with no significant differences in the occurrence per dataset. %\TODO{REPHRASE? with a range from 22 to 27 mentions}),
This included the number of rows, columns, or instances (such as the number of countries in the data). For instance: \textit{"Size: $468$ rows by $32$ columns (incl. headers)"} or \textit{"information on $101,171$ earthquakes"}. Additionally, some summaries include the number of possible values which can be expected in a specific column, such as in this example for the header "hair": \textit{"HAIR - TEXT - $23$ hair colours plus bald and no hair"}.

Possible values in a column were mentioned explicitly by 56.6\% of all participants, most often in connection to D2. We assume that is because this dataset has a number of columns in which the range of values is limited. For instance headers referring to eye or hair colour or gender which have a limited number of possible entries:

\begin{quote}
%\vspace{-5px}
(P20) The dataset also characterises whether the characters are good, bad, or neutral.
%\vspace{-5px}
\end{quote}

When there is a greater number of possible values these were presented through ranges or examples or by defining data types or other constraints for a column.

\begin{quote}
%\vspace{-3px}
(P21) ID: Identity is secret/public/etc. ALIGN: Good/bad/neutral/-etc. EYE: Character's eye colour HAIR: Character's hair colour
%\vspace{-3px}
\end{quote}

It is likely that the number of explicit mentions of possible values is under representing the importance of this category: As the participants were describing the dataset for someone else and in natural language we would assume that if the summary specifies e.g. "age", there is no need to further  explain this column presents the value type numbers as this would automatically be inferred, such as in a conversation between people. E.g. if there is a header called "age", we expect the value type to be numerical.

\section{Discussion}
\label{sec:discussion}

We discuss the identified summary attributes, the results of the diary study
and how these insights can inform the design of automatic summary creation. We compare our findings to existing metadata guidelines and detail the implications our results have on defining user centred dataset summaries.
We then present a template for user centred dataset summaries which can be incorporated into data portals, used by data publishers and inform the development of automatic summarisation approaches. We conclude by discussing where we see the role of textual summaries, together with metadata, in the data discovery process.

\subsection{Summaries attributes}
\begin{figure*} [h]
\centering
\begin{footnotesize}
\includegraphics[width=100mm]{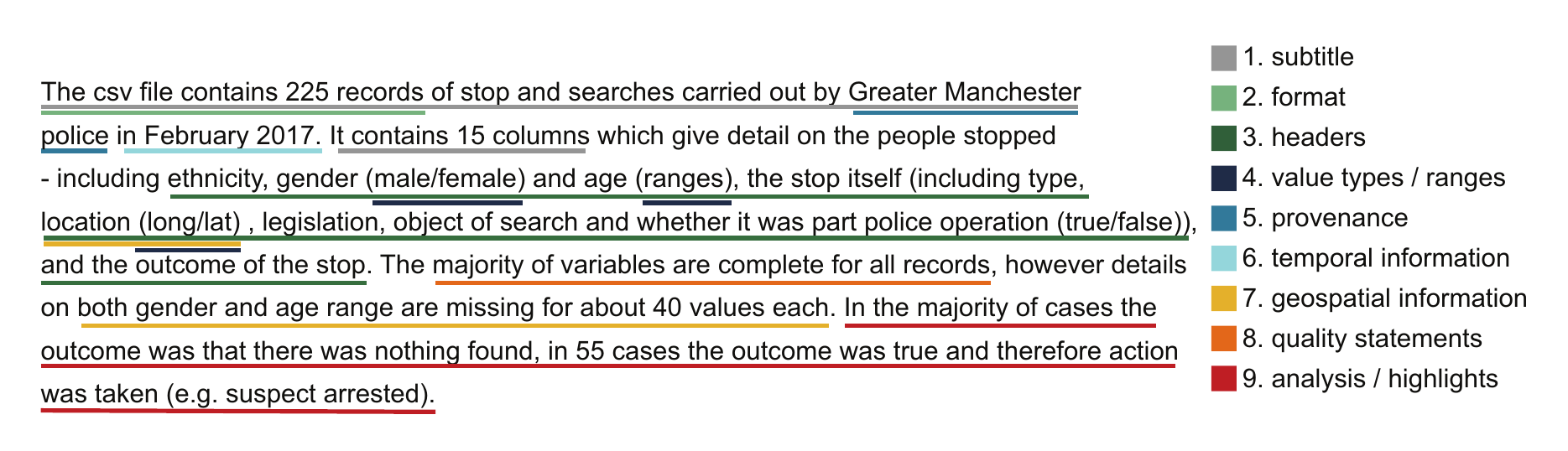}
\end{footnotesize}
\caption{Example of an annotated dataset summary}
\label{fig:examplesummary}
%\vspace{-2mm}
\end{figure*}

%\noindent \textbf{Summaries attributes}
We identified features that people consider important when trying to select a dataset (RQ1), and when trying to convey a dataset to others (RQ2), as can be seen in Table \ref{tab:comparison}.
Our findings address a gap in literature, relevant in the context of data publishing, search and sharing. We were able to see common structures and isolate different attributes that the summaries were made of (RQ2), as can be seen in Figure \ref{fig:examplesummary}.
Summaries for the same dataset, created by different participants shared common attributes. We found a number of attributes tend to be less dependent on the underlying datasets, such as subtitle, format, headers and quality; whereas others tend to vary more depending on the data. Our findings allowed us to determine the composition and feasibility of general purpose dataset summaries, written solely based on the content of the dataset, without any further context.

Our findings suggests a range of datasets characteristics which people consider important when engaging with unfamiliar datasets. This analysis allows us to devise a template for the creation of text representations of datasets which is detailed in \ref{sec:template}.
Some of the attributes could be generated automatically, while others would still require manual input, for example from the dataset creator or from other users. We saw that all dataset summaries, as expected, explicitly describe the scope of the content in the dataset. Extracting content features directly from the dataset, and representing them as text is still subject of research, in particular in the context of extractive dataset summarisation \citep{ferreira2013assessing} or semantic labeling of numerical data \citep{Pham2016}. Our findings can inform the design of these methods by suggesting parts of a dataset that matter in human data engagement.

In the same time, our analysis shows that most summaries also cover information that goes beyond content-related aspects, including groupings of headers into meaningful categories, the identification of key columns, and in some cases also the relationship between these and other columns in the dataset. These areas should be taken into account by data publishers when organising and documenting their data, and by designers of data exploration tools. For example, tools could highlight key columns and their relationships, or display structure overlaps that group headings in a relevant way. Furthermore, our summaries contained quality statements, some of which are complex as they refer to the potential context or use cases of the dataset; or an expression of uncertainty. We therefore conclude that purely extractive approaches will unlikely be able to produce useful text summaries of datasets that meet people's information needs.

While abstractive approaches to automatically generate summaries exist, we believe that the levels of abstraction and grouping needed for the creation of meaningful textual representations of data are not yet being realised. To be truly useful, a summary needs to be a combination of extractable features, combined with contextual information, human judgement, and creativity. This applies to selecting the right content to consider, as well as to representing this content in a meaningful way.

Comparing summaries created in a lab setting to those created in a crowdsourcing experiment gave us an understanding of the level of expertise or the \textit{closeness} to the data that is needed to write a meaningful summary. It further gives insights into the feasibility of crowdsourcing as a potential method for dataset summary generation. We found dataset summaries can be produced using crowdsourcing, however, to fully reproduce summaries as they were created in the lab experiment crowdworkers could benefit from additional guidance, such as a template to support the summary writing process. We believe such a template would equally facilitate data publishers to write a comprehensive and meaningful summary and is necessary for the development of automated dataset summarisation approaches.

Without this research, researchers and developers creating summaries would focus on obvious items such as column headers. This work demonstrates the importance of other aspects such as the grouping of headers, value types and ranges, information about data quality or usage suggestions - all attributes not commonly included in metadata. This highlights the difficult areas in fully automated approaches to summary creation. Understanding which attributes are considered important when selecting and describing datasets can focus future research efforts to deliver value to users. It can also be used to inform benchmark design for automated summary creation research.

%\noindent \textbf{Comparison to metadata standards and data search
\subsection{Comparison to metadata standards and data search diaries} Table \ref{tab:comparison} shows a comparison between the results of the summary creation study, the outcomes of the analysis of data search diaries and current metadata standards.
We can see that the attributes \textit{basic stats, quality statements, patterns/trends and usage} are currently not represented in either of the two metadata schemas we discuss. Further differences include the grouping of column headers in meaningful semantic categories, the identification of a key column, and the importance of value types for the main columns.

We saw that many summaries, as well as the diary data suggest the usefulness of basic statistics about the dataset, such as the number of rows and columns, but also information on the possible values or ranges of important columns. These are potentially easy to extract from a dataset but are not usually captured in standard metadata. In terms of geospatial and temporal attributes the main difference concerns the granularity of the information.
Quality statements, initial analysis of the dataset content (patterns and trends) and ideas for usage are those attributes which are potentially complex to create but can be of great value in the selection process of datasets \citep{Laura}. We believe that both provenance and methodology are under represented in the summaries due to the nature of the task and experiment design. Our work focuses on attributes people find important when selecting and describing datasets. However, whether the attributes should be represented in textual summaries or as structured metadata would be an interesting direction of future research.

\def\arraystretch{1.3}
\begin{table}[H] %\sffamily
\centering
%\normalsize %default font size
%\small
%\footnotesize
\scriptsize
%\tiny %Smallest font size
%\setlength{\tabcolsep}{10pt} % Default value: 6pt
%\renewcommand{\arraystretch}{1.5} % Default value: 1
\setlength{\tabcolsep}{1em}
\begin{tabular}{p{2.7cm}|p{4cm}|p{4.1cm}|p{3.7cm}}
%\hline \hline
 %\multirow{2}{*}{\textbf{SUMMARIES}} & \multirow{2}{*}{\textbf{DIARY}} & \multirow{2}{*}{\textbf{SCHEMA / DCAT}} \\ [-0.2ex]
% & &  \\[-0.2ex] \hline %\hline
%\rowcolor[HTML]{b1cbe2}
\hline %\hline
\rowcolor[HTML]{b1cbe2} \textbf{Category} & \textbf{Summaries} & \textbf{Diary} & \textbf{Schema and DCAT} \Theader  \\ \hline
 \textbf{Format} and file related info &
file format, size of the file, personal data, last updated, license, unique identifiers  & file format, api, access, unique identifiers, language, size  & S: file format, license, identifier, url, D: size (bytes), format, identifier, language  \\ \hline

 \textbf{Provenance} &
provenance: publisher, publishing organisation, \newline temporal provenance: publishing date, last update, time of data collection \newline geospatial provenance & publishing org (authoritative, reliable source), funding organisation (bias, independent source), original purpose (context)  & S: author, contributor, producer, publisher, creator, editor, provider, source organisation \newline  D: contact point; publisher, landing page,  sponsor, funder \\ \hline

\textbf{Subtitle} &
 high-level one phrase summary &  title &  S: main entity, about, headline  \newline D: theme, concept, keyword, title  \\ \hline

\textbf{Headers} and \textbf{Groupings} &
%\rowcolor[HTML]{e8e8e8}   \rule{0pt}{13pt} \textbf{HEADERS}  & &\\ %\hline
headers, selection and grouping of headers (+explanation), key columns & headers, attributes/values and their meaning, value types (documentation) & S: variables measured   \\ \hline

 \textbf{Geographical} &
geospatial scope (+level of granularity)  & location of publishing organisation, geospatial coverage (level of granularity) &  S:  location created, spatial coverage, content location \newline D: spatial coverage     \\  \hline
 \textbf{Temporal} &
temporal coverage  (+level of granularity),  & temporal scope, level of granularity, time of data collection (including time of the year), temporal provenance (time of publishing, up-to-date, maintained) & S: temporal coverage, content reference time, date created, date modified, date published \newline  D: temporal, temporal coverage, release date,  update date, frequency of publishing  \\ \hline

\textbf{Quality} & \textit{quality dimensions:} & & - \\
 &
completeness \newline
consistency in formatting \newline
understandability (headers, acronyms, abbreviations) \newline
representativeness, coverage
&
completeness, accuracy \newline
consistency in formatting, cleanliness \newline
understandability, clear provenance and authoritativeness of source \newline
 -
&   \\  \hline
%  &   &    \\
%& consistency in formatting &  &    \\
%& understandability (headers, acronyms, abbreviations) &  &    \\
%&   & - &   \\ \hline

\textbf{Basic statistics} &
ranges per column (possible values per column), \newline counts of rows and columns,  size \newline possible value ranges and data types & units of measurement, upper/lower bounds to estimates,unique values for a column, comprehensiveness, range and variation, number of rows and columns & -   \\ \hline

 \textbf{Patterns and Trends} &
analysis of the dataset content (patterns, trends, highlights) & - &  -  \\ \hline

 \textbf{Usage} &
  ideas for usage & reasons not to use the dataset & -   \\ \hline

Methodology &
 - & methods, control group, randomised trial, number of contributors, confidence intervals, sample and consideration of influencing factors, bias, sample time  & S: measurement technique, variables measured   \\ \hline
%\multicolumn{4}{c}{\rule{0pt}{13pt} \textbf{LANGUAGE}} \\ \hline
 %& language &  & language \\ \hline
\end{tabular}
\caption{\footnotesize Comparison of summary attributes to data-search diary and metadata standards. Summary = results from this study; Diary = Analysis of selection criteria in a data-search diary; Schema (S) = {http://schema.org/Dataset} DCAT (D) = {https://www.w3.org/TR/vocab-dcat/} - Attributes "description" excluded}
\label{tab:comparison}
\end{table}

\subsection{Making better summaries}

Prior work has identified dataset relevance, usability and quality as critical to dataset search \citep{Laura}.
Relevance can be determined by having insights into what the dataset contains, and by analysing the data. Usability can be judged from the descriptive information in the summaries (such as format, basic stats, license, etc.). The quality and uncertainty statements expressed in the summaries deliver an assessment of dataset quality.

Individual attributes of the summaries could be generated using existing approaches, for instance from database summarisation methods some of which generalise column content into higher level categories, ideally describing the content in the column \citep{DBLP:conf/vldb/Saint-PaulRM05}. Other approaches have tried to automatically identify the key column of a dataset \citep{ermilov_taipan_2016,venetis_recovering_2011}.

\noindent \textbf{Granular temporal and location descriptions.}
%\subsection{Granular temporal and location descriptions}
Among the results that confirmed existing best practices and standards were the prevalence of time and location in characterising datasets. These are commonly covered by existing metadata formats.\footnote{\url{http://schema.org/Dataset}} Our study has revealed a multitude of granularities in connection to these features, which are less well supported.
The level of granularity of temporal or geospatial features of a dataset is crucial to understand its usefulness of a dataset for a particular task. This is reflected in the number of indications of these attributes in the summaries. Based on the results of this study we believe summaries should support users to determine whether a dataset has appropriate levels of aggregation for a given task.

\noindent \textbf{Standard representations of quality and uncertainty.}
%Data Quality Assessment Pipino:2002:DQA:505248.506010
Quality statements in the summaries included judgements on completeness, as well as assumed comprehensiveness of the data, errors and precision. Uncertainty statements referred to the meaning of concepts or values in the dataset (commonly including abbreviations and specialised terms) - which confirms findings in \citep{Laura} as well as unclear temporal or geographical scope of the data. Such statements illustrate the potential impact that good textual summaries and documentation can have for data users. W3C guidelines include
completeness and availability as quality-related measures\footnote{\url{https://www.w3.org/TR/dwbp/}}.
Our study shows that, especially in the more in-depth lab summaries, statements expressing uncertainty or sanctioning the quality of a dataset are very common. There is a body of research discussing how to best communicate uncertainty in visual representations of data, for instance \citep{Boukhelifa:2017:DWC:3025453.3025738,Kay:2016:MBU:2858036.2858558,simianu2016understanding}. Understanding how to communicate uncertainty in textual representations of data, and furthermore, how this type of information impacts on the decisions of subsequent data users and on the ways they process the data, is comparatively less explored. Furthermore, previous research with data professionals has suggested that assessing data quality plays a role in selecting a dataset out of a pool of search results; studies such as \citep{DBLP:journals/corr/GregoryGCSW17,Laura,DBLP:journals/jmis/WangS96} have discussed the task-dependent and complex nature of quality. We assume that creating a more standardised way of representing uncertainty around datasets would be beneficial from a user perspective; related literature indicates that communicating uncertainty improves decision making and increases trust in everyday contexts \citep{joslyn2013decisions,kay2013there}.

\textbf{Summary length.} One open question in the context of summary creation is the optimal length of a general purpose dataset summary. Regarding the effect of summary length - our study showed that the longer summaries produced in the lab experiment contained more qualitative statements which not only describe the data but judge the dataset for further reuse. This is not to say that, in all cases, the longer a summary, the better its quality.
It is important to consider the likelihood that there is an optimal summary length, and surpassing this causes quality to decrease as the key elements of the summary become less accessible - which is an interesting area for future work. Determining snippet length in web search has been subject of numerous studies, for instance \citep{DBLP:conf/chi/CutrellG07, DBLP:conf/coling/HeDN12, DBLP:conf/sigir/MaxwellAM17} which generally suggest summary length influences relevance judgements by users. However, in this work we focus on summary content and not on summary length.
%In terms of presentation modes of the summaries,
A small scale user study by \citet{DBLP:conf/adcs/AuTJ16} tested the presentation mode for automatically created query-biased summaries from structured data and suggests a preference for non-textual summaries. However, their textual summaries are limited in scope and fluency and are so not comparable to what we refer to as meaningful summaries in the context of this work.

%\noindent \textbf{Dataset summary template}
\subsection{Dataset summary template}
\label{sec:template}
Studies on text summarisation found that people create better summaries when they are given an outline or a narrative structure that serves as a template, as opposed to having to create text from scratch \citep{borromeo2017crowdsourcing, kim2016storia}. Based on our findings, we propose such a template for text-centric data summaries. If used it could improve current practices for manually written summaries and potentially inform automatic data-to-text approaches.%\newline

\begin{multicols}{2}
\begin{enumerate}[leftmargin=*]
\scriptsize
\setlength\itemsep{-0.5em}
\item \textbf{How would you describe the dataset in one sentence?}
\item \textbf{What does the dataset look like?} File format, number of rows and columns, machine readability
\item \textbf{What are the headers?} Can you group them in a sensible way? Is there a key column?
\item \textbf{What are the value types and value ranges for the most important headers?} Words/numbers/dates and their possible ranges
\item \textbf{Where is the data from?} When was the data collected/published/updated? Where was the data published and by whom?
\item \textbf{In what way does the dataset mention time?} What timeframes are covered by the data, what do they refer to and what is the level of detail they are reported in? (E.g. years/day/time/hours etc.)
\item \textbf{In what way does the dataset mention location?} What geographical areas does the data refer to? To what level of detail is the area or location reported? (E.g. latitude/longitude, streetname, city, county, country etc.)
\item \textbf{Is there anything unclear about the data, or do you have reason to doubt the quality?} How complete is the data (are there missing values)? Are all column names self explanatory? What do missing values mean?
\item \textbf{Is there anything that you would like to point out or analyse in more detail?} Particular trends or patterns in the data?
\end{enumerate}
\end{multicols}

\begin{comment}
\begin{enumerate}[leftmargin=*]
\scriptsize
\setlength\itemsep{-0.5em}
\item \textbf{How would you describe the dataset in one sentence?}
\item \textbf{What does the dataset look like?} File format, number of rows and columns, machine readability
\item \textbf{What are the headers?} Can you group them in a sensible way? Is there a key column?
\item \textbf{What are the value types and value ranges for the most important headers?} Words/numbers/dates and their possible ranges
\item \textbf{Where is the data from?} When was the data collected/published/updated? Where was the data published and by whom?
\item \textbf{In what way does the dataset mention time?} What timeframes are mentioned, what do they refer to and what is the level of detail they are reported in? (E.g. years/day/time/hours etc.)
\item \textbf{In what way does the dataset mention location?} What areas does the data refer to? To what level of detail is location reported? (E.g. latitude/longitude, streetname, city, county, country etc.)
\item \textbf{Is there anything unclear about the data, or do you have reason to doubt the quality?} Are all column names self explanatory? How complete is the data?
\item \textbf{Is there anything that you would like to point out or analyse in more detail?} Particular trends in the data?
\end{enumerate}
\end{comment}

Our findings showed a dependency of attributes on the dataset content, mostly for temporal information, meaningful groupings of headers, provenance, basic stats and geospatial information (which might be an exception, as explained in the findings).
Hence we suggest number 1-4 to be required, as they are generic attributes describing datasets. Number 4, a datasets provenance, is usually provided in standard metatdata. The template questions number 5-9 are considered to be optional in the summary, as they not necessarily applicable for all datasets.

The template focuses on attributes that can be inferred from the dataset itself, or on information that is commonly available in metadata, such as provenance. For instance, we do not include uncertainty about the dataset as a template question as the summaries have shown that uncertainty statements can refer to any of the categories of the template and is inherently dependent on the user.

%\TODO{expl why uncertainty not included in the template, can refer to any of the categories of the template, to generic and subjective}

%\noindent \textbf{From summaries to metadata}
\subsection{From summaries to metadata} While our focus was on text summaries, the themes we have identified can inform the design of more structured representations of datasets, in particular metadata schemas as a primary form for automatically discovering, harvesting, and integrating datasets. Like any other descriptor, metadata is goal-driven, it is shaped by the type of data represented, but also by its intended use \citep{greenberg2010metadata}. Text summaries of data can be seen as metadata for consumption by people. They are meant to help people judge the relevance of a dataset in a given context. Structured metadata, commonly in form of attribute value pairs, is potentially useful in this process as well; in fact, in the absence of textual summaries, people use whatever metadata they can find to decide whether to consider a dataset further. However, metadata records are primarily for machine consumption; they define a set of allowed attributes, use controlled vocabularies to express values of attributes, and are constrained in their expression by the need to be processable by different types of algorithms. This contrast is what makes text summaries of datasets so relevant for HCI - these are often the first "point of interaction" between a user and a dataset \citep{Laura}. Beyond that, we nevertheless believe that some of their most common content and structural patterns can inform the design of automatic metadata extraction methods, which in turn could improve dataset search, ranking, and exploration. For instance, knowing that the number of rows and headers in a datasets help users to determine a dataset's relevance, means these comparably easily extractable attributes could be included in automatic metadata extraction methods.
Our results point to a number of attributes that could easily be extracted, but for which there is no standard form of reporting in general-purpose metadata schema. These include descriptive attributes such as the mentioned numbers of rows and headers, possible value types and ranges, as well as different levels of granularity of temporal or geospatial information. A one-sentence summary, which has also been found to be useful by \citet{yu2007choosing} in a study on expert summarisation of time series data, or meaningful semantic groups of headers are more complex to create.
Further complex features include the variety of elements which describe quality judgements and uncertainty connected to the data; and the identification of a key column.

\section{Limitations}
\label{sec:limitations}
The dataset searching diaries consisted of set questions that make the person writing the response think about partially subconscious selection process in an abstract way and requires them to articulate their information needs. Although that is a potentially complex task we our findings suggest that participants expressed real information needs and the results generally overlapped with those in \citep{Laura}. However, observational studies could be done in future work to confirm or complement these findings through different methods.

There are several confounding factors in the task of summary generation,
due to the complexity of the task, which was also discussed in previous research on textual summary creation \citep{Bernstein:2015:SWP:2808213.2791285}. The overarching aim of this study was to gain an understanding of peoples' conceptualisation of data, within the boundaries of this task (instructions, environment, time constraints). We did not specify the desired output in our lab experiment in terms of structure, style, choice of features and type of language, as we wanted to see what type of summaries people produce without guidance.
%We assume that our choice of crowdsourcing as a method to evaluate the quality of the summaries resulted in lower interrater reliability scores then for instance ratings in a lab experiment would have\TODO{why?}.

The study was carried out using files in CSV format; while we assume that elements of the summaries, particularly the high-level information extraction, would likely remain the same for all structured or semi-structured data, the description of the structure and representation (such as the number of rows, headings, etc.) of a dataset using a different format might vary.
We found the particular datasets influenced the composition of the summaries in some instances, such as quality statements. geospatial attributes and provenance. However, despite these differences, we believe that there were sufficient commonalities in the summaries, both between datasets and the methods used, to derive recommendations and identify directions for further improvement. We acknowledge that whenever we are attempting to develop a more standardised way of documentation in a domain as open as data search, guidelines will not fit every scenario to the same extent. This is why we chose a variety of dimensions in the dataset sample, aiming for a template that covers different types of data and could potentially be extended for more specific requirements.

Participants in the lab experiment were data literate and used data in their work, but did not necessarily classify themselves as data professionals. As a result, they may not have been aware of additional needs of data professionals, such as information on licensing or formatting that might have been mentioned, had they more specialised knowledge. Furthermore, we suspect personal data would in reality play a bigger role in a different study, for relevant datasets. More research would be needed to understand how summaries would change when sensitive information is present.
We used publicly available datasets that are not known to be popular, though we cannot be certain that none of our participants were familiar with the datasets. However, literature on text summarisation found that prior knowledge did not have significant effects on written summarisation performances \citep{yu2009shifting}.
While we believe that there is intrinsic value in textual summaries of datasets - as they cannot only be used to inform selection by users, but could also be useful in search - we do not test the best representation of summary content in this work. Further studies are needed to determine optimal presentation modes of summary content for user interaction in a dataset selection activity.

\section{Conclusion and Future Work}
\label{sec:conclusion}

With the overabundance of structured data, techniques to represent it in a condensed form are becoming more important. Text summaries serve this function and they have the potential to make data on the web more user friendly and accessible.
We contribute to a better understanding of human-data interaction, identifying attributes that people consider important when they are describing a dataset. We have shown that text summaries in our study are laid out according to common structures; contain four main information types; and cover a set of dataset features (RQ1). This enables us to better define evaluation criteria for textual summaries of datasets; gives insights into selection criteria in dataset search; and can potentially inform metadata standards.

We conclude that our results are consistent enough between different participants and between different types of datasets to assume their generalisability for our scenario (RQ2). We found general overlap between the information needs expressed in the data-search diaries (RQ1) and in the summaries created as a result of this study.
Based on a subset of attributes, we found that summaries of data practitioners have a higher prevalence of provenance, quality statements and usage ideas as well as a slightly more geospatial information. We also found that a number of attributes depend more on the dataset than others and which could influence the application of the dataset summary template.

Our results further suggest that crowdsourcing could be applied for large-scale dataset summarisation, however the validity would need to be studied in more depth. This study gives first insights into the feasibility of such an approach. Furthermore, when indexing dataset content to support search, we need to make a selection of important attributes based on what people search for and choose to summarise about a dataset.

The attributes mentioned in the summaries could also indicate those that are useful in search, which, if validated in future work, could increase the discoverability of data on the web. Web search functionalities are tailored to textual sources, therefore having a textual summary containing meaningful content on the dataset could potentially allow general web search engines to index data sources in a similar way as web pages.

This work could be extended in a number of directions. We aim to evaluate the perceived usefulness of summaries created according to the proposed template as a next step. Follow-up studies could include crowdworkers iterating on the summaries created by the template, which has been proven useful for image descriptions and text shortening \citep{Bernstein:2015:SWP:2808213.2791285,Little:2010:EIP:1837885.1837907}. Additional work could be carried out on refining a semi-automatic approach to generating summaries, using the template by prompting crowd workers to extract these elements from datasets. This may also have the side-effect of producing higher quality descriptions overall, simply by providing more structure to the task and clearer examples and guidance to the crowd workers, as well as validation and training. There is a large body of research aiming to understand visual representations of data for different contexts. Similarly we believe that we need to further examine textual representations for data in much more detail to understand how to tailor them to specific users and their contexts. Similarly, approaches to generate query-biased summaries, such as shown by \citep{DBLP:conf/adcs/AuTJ16} to generate task dependent summaries are an interesting area for further research.
%\TODO{different summaries for different users, query biased summaries}
\\
\section{Acknowledgements}
This project is supported by the European Union's Horizon $2020$ research and innovation programme under the Marie Sk\l{}odowska-Curie grant agreement No 642795. We thank our participants for taking part in this study.

%% The Appendices part is started with the command \appendix;
%% appendix sections are then done as normal sections
%% \appendix

%% \section{}
%% \label{}

%% If you have bibdatabase file and want bibtex to generate the
%% bibitems, please use
%%
\section*{References}
\bibliographystyle{elsarticle-harv}
\bibliography{bib}

%% else use the following coding to input the bibitems directly in the
%% TeX file.

%\begin{thebibliography}{00}

%% \bibitem[Author(year)]{label}
%% Text of bibliographic item

%\bibitem[ ()]{}

%\end{thebibliography}
\end{document}